 \documentclass[twocolumn,showpacs]{revtex4}

\usepackage[latin1]{inputenc}
\usepackage{graphicx}%
\usepackage{dcolumn}
\usepackage{amsmath}

\makeatletter
\def\btt#1{\texttt{\@backslashchar#1}}%
\DeclareRobustCommand\bblash{\btt{\@backslashchar}}%
\makeatother

\begin{document}

\title{Failure of classical traffic and transportation theory:  
 The maximization of the  network throughput  maintaining 
 free flow conditions   in  network
}

\mark{}

\author{   
Boris S. Kerner$^1$}

 \affiliation{$^1$
Physics of Transport and Traffic, University Duisburg-Essen,
47048 Duisburg, Germany}



\begin{abstract} 
We have revealed  
general physical conditions
 for the {\it maximization} of the network throughput at 
which free flow conditions are ensured, i.e., traffic breakdown cannot occur
 in the whole traffic or transportation network.
 A physical   measure of the network  -- {\it  network
capacity} is introduced that characterizes general features of the network with respect to the
 maximization of the network throughput.   
	The   network capacity allows us also to make  a general   proof of the deterioration of
	traffic system occurring when dynamic traffic assignment is performed in a network based on the classical
	Wardrop' user equilibrium (UE) and system optimum (SO) equilibrium.
\end{abstract}

 \maketitle

 \section{Introduction \label{Int}}
 
To find traffic optima  and organize  dynamic traffic assignment and
 control in traffic and transportation networks,
a huge number of traffic theories   
have been introduced 	(see, e.g.,~\cite{Wardrop,Sheffi1984,BellIida19897,Peeta2001A,Mahmassani2001A,Rakha2009,DaganzoSheffi1977,Merchant1978A,Merchant1978B,Bell1992A,Bell2002A,Mahmassani1987A,Friesz2013A,Ben-Akiva2015A,Wahle,Davis1,Davis2,Wang2005A,Claes2011A,Donga2010A,Sopasakis2006,Hauck2014,Sun2014A}
and references there). Some of  traffic models developed recently  are devoted to
the development of optimal 
dynamic feedback strategies in the networks (e.g.,~\cite{Wahle,Davis1,Davis2,Wang2005A,Claes2011A,Donga2010A})
and search algorithms based on stochastic processes which find local optima with asymmetric
 look-ahead potentials (e.g.,~\cite{Sopasakis2006,Hauck2014,Sun2014A}). 

The most famous approach to an analysis of traffic and transportation networks is based on the Wardrop's  
user equilibrium (UE) and system optimum (SO) equilibrium introduced 
 in 1952~\cite{Wardrop}.
	{\it Wardrop's UE}: traffic on a network distributes itself in such a way that the
  travel times on all routes used from any origin to any destination are equal, while all unused routes have equal or greater travel   times. 
     {\it Wardrop's SO}:  the network-wide travel time should be a minimum. 
 Wardrop's principles reflect either the wish of  drivers to reach their destinations as soon as
   possible (UE) or the wish of network operators to reach the minimum network-wide travel time (SO).
	During last 50 years based on   the Wardrop's  
 equilibria a huge number of theoretical works to 
 dynamic traffic assignment and control in the networks have been made by several generations
of traffic and transportation researchers   (see, e.g.,~\cite{Sheffi1984,BellIida19897,Peeta2001A,Mahmassani2001A,Rakha2009,DaganzoSheffi1977,Merchant1978A,Merchant1978B,Bell1992A,Bell2002A,Mahmassani1987A,Friesz2013A,Ben-Akiva2015A} and references there). 
	
	In particular, in accordance with
	the Wardrop's   equilibria,   
travel times or/and other travel costs on network links
 are considered   to be self-evident
 traffic characteristics for
 objective functions used for optimization 
transportation problems~\cite{Sheffi1984,BellIida19897,Peeta2001A,Mahmassani2001A,Rakha2009,DaganzoSheffi1977,Merchant1978A,Merchant1978B,Bell1992A,Bell2002A,Mahmassani1987A,Friesz2013A,Ben-Akiva2015A}.
The main aim of associated classical approaches is to minimize
	travel times or/and other travel costs  in a traffic or transportation
	network
	(see, e.g.,~\cite{Sheffi1984,BellIida19897,Peeta2001A,Mahmassani2001A,Rakha2009,DaganzoSheffi1977,Merchant1978A,Merchant1978B,Bell1992A,Bell2002A,Mahmassani1987A,Friesz2013A,Ben-Akiva2015A,Wahle,Davis1,Davis2}). 
The classical traffic and transportation
 theories based on Wardrop's equilibria have made a great impact on the understanding of
many traffic phenomena. However, network optimization approaches based on these theories
have failed by their applications in the real world (see explanations of these critical statements and references in critical reviews~\cite{Kerner_Review,MiniReview2}). Even several decades of a very intensive effort to improve and validate
network optimization models have had no success. Indeed, there can be found no examples where on-line implementations
of the network optimization models based on these theories could reduce congestion in real traffic
and transportation networks.

As explained in recent reviews~\cite{Kerner_Review,MiniReview2},  
 assumptions about traffic features used for the development of
 classical  traffic flow models
  are not consistent with the empirical nature of traffic breakdown at
	network bottlenecks. The empirical
nature of traffic breakdown explained in
  three-phase traffic   theory~\cite{KernerBook,KernerBook2}
is as follows: Traffic breakdown is a phase transition from
free flow (F) to synchronized flow (S) that occurs
in a {\it metastable state of free flow} at a network bottleneck. 
Each of   network bottlenecks    is characterized by
a minimum   capacity  $C_{\rm min}$ that separates
stable and metastable states  of free flow at the bottleneck~\cite{KernerBook,KernerBook2}:  
At
\begin{equation}
  q_{\rm sum} < C_{\rm min}, 
    \label{C_min_q_in} 
    \end{equation}
free flow is stable at the bottleneck,
where $q_{\rm sum}$ is the flow rate in free flow at 
the bottleneck~\cite{Kerner_Review,KernerBook}. Therefore, 
under condition (\ref{C_min_q_in}) no traffic breakdown can occur at the bottleneck. 
Contrarily, at
\begin{equation}
  q_{\rm sum} \geq C_{\rm min} 
    \label{C_min_q_in_larger} 
    \end{equation}
	free flow is metastable with respect to traffic breakdown
	at the bottleneck. Therefore, under condition (\ref{C_min_q_in_larger})	traffic breakdown can occur at the bottleneck.

In this paper, we reveal
	an approach to the maximization of the network throughput
	ensuring free flow conditions at which
	traffic breakdown cannot occur in the whole network. We call this approach
	{\it network throughput maximization approach}.
	We introduce  
	a physical  measure of a traffic or transportation network called {\it network capacity}.
	The network capacity
  characterizes general features of the network with respect to the
 maximization of the network throughput at which free flow conditions are ensured in the networks.
	  One of the consequences of the general analysis of traffic and transportation networks
	made in the article	is  a general   proof of the deterioration of
	traffic system occurring when dynamic traffic assignment is performed in a network based on the  
	Wardrop'  equilibria. This can explain the failure 
  of the application of the Wardrop'  equilibria
	for the prevention of traffic congestion in real traffic and transportation networks.
  
 The article is organized as follows. In Sec.~\ref{GenBM_S},
	we introduce the network throughput maximization approach for the
prevention of breakdown in   networks at the maximum network throughput as well as the network capacity.   In discussion Sec.~\ref{Dis_S}   we   make a general analysis of 
the deterioration of the traffic system	through the use of Wardrop's equilibria,
  illustrate this
	general analysis by numerical simulations,
  discuss   how   appropriate information to the drivers can be given to organize
the maximization of the network throughput   as well as
consider a connection between 
the network throughput maximization approach introduced in this article and
  the breakdown minimization (BM) principle of Ref.~\cite{BM_BM,BM_BM2}.

\section{The maximization of the  network throughput ensuring
 free flow conditions   in  network \label{GenBM_S}}

\subsection{Network model}

In   known real field (empirical) traffic data, 
traffic breakdowns occur usually at the same road locations of a traffic network called network
 	bottlenecks (e.g.,~\cite{May1990,ElefteriadouBook2014,Reviews2,KernerBook,KernerBook2}).
		Network bottlenecks are caused, for example, by
	on- and off-ramps, road gradients, road-works, a decrease 
	in the number of road lines (in the flow direction),  traffic signals in city traffic, etc.
	A bottleneck introduces a speed disturbance localized
	in a neighborhood of the bottleneck. As a result,
	in the empirical data 
	at the same flow rate the probability of traffic breakdown at a bottleneck on a network link 
	is considerably
	larger than on the link outside the bottleneck.

We consider a traffic or transportation network with
 $N$ network bottlenecks, where  $N>1$. We assume that there are
$M$ network links  (where $M>1$) for which the inflow rates $q_{m}$, $m=1,2,\ldots,M$
can be adjusted;
$q_{m}$ is the link inflow rate
for a link with index $m$. 
At the network boundaries, there are $I$ links  
for the network inflow rates 
 $q^{\rm (o)}_{i}(t)$, $i=1,2,\ldots,I$
(called $\lq\lq$origins", 
for short $\rm {O}_{i}$, $i=1,2,\ldots,I$), where $I\geq 1$. 
At the network boundaries, there are also $J$ links  for the network outflow rates  
$q^{\rm (d)}_{j}(t)$, $j=1,2,\ldots,J$ 
  (called $\lq\lq$destinations", for short $\rm {D}_{j}$, $j=1,2,\ldots,J$), where $J\geq 1$.
	The network inflow rates $q^{\rm (o)}_{i}(t)$ and the total network inflow rate $Q(t)$ 
	are determined by   the network inflow rates $q^{\rm (o)}_{ij}(t)$ of vehicles moving
	from origin $\rm {O}_{i}$ to destination $\rm {D}_{j}$ 
(called as origin-destination pair $\rm {O}_{i}$--$\rm {D}_{j}$ of the network)
through the well-known formulas
	\begin{equation}
	q^{\rm (o)}_{i}(t)=\sum^{J}_{j=1}{q^{\rm (o)}_{ij}(t)}, \quad
	Q(t)=\sum^{I}_{i=1}{q^{\rm (o)}_{i}(t)}.
	\label{OD_F}
	\end{equation}
In this article, 
the network inflow rates 
 $q^{\rm (o)}_{ij}(t)$ 
(origin-destination matrix) are assumed to 
be known time-functions.
 Each of the network bottlenecks  $k=1,2,\ldots,N$ is characterized by
a minimum highway capacity $C^{(k)}_{\rm min}=C^{(k)}_{\rm min}(\alpha_{k}, {\bf R}_{k})$,
  $\alpha_{k}$   
is the set of control parameters  
of    bottleneck    $k$~\cite{ControlParemeters};  
${\bf R}_{k}$  is a matrix
of percentages of vehicles with different vehicle (and/or driver) characteristics
that takes into account that dynamic assignment
is possible to  perform  individually for each of the vehicles~\cite{DiffVehicles};
the flow rate in free flow at 
bottleneck $k$ will be denoted by $q^{(k)}_{\rm sum}$.

In contrast with   classical traffic and transportation  
theories (see, e.g.,~\cite{Sheffi1984,BellIida19897,Peeta2001A,Mahmassani2001A,Rakha2009,DaganzoSheffi1977,Merchant1978A,Merchant1978B,Bell1992A,Bell2002A,Mahmassani1987A,Friesz2013A,Ben-Akiva2015A}), our approach for the maximization of the network throughput 
does not include travel times (or other $\lq\lq$travel costs") on  different network routes.
To avoid the use of   network routes with 
considerably longer travel times   in comparison with
 the
 shortest routes,  
 the network throughput maximization approach to dynamic traffic assignment and network control is applied for some $\lq\lq$alternative network routes (paths)" (for short,
alternative  routes) {\it only}. This means that 
there is a constrain $\lq\lq$alternative network routes (paths)" by the application of the network throughput maximization approach.

We define the
constrain $\lq\lq$alternative network routes (paths)" as follows. The alternative routes   are
 possible  different routes from origin $\rm {O}_{i}$ to 
destination $\rm {D}_{j}$ (where $i=1,2,\ldots,I$ and $j=1,2,\ldots,J$)
	for which
 the maximum difference between travel times in free flow conditions   does not exceed
 a {\it given value} denoted by $\Delta T_{ij}$ (values $\Delta T_{ij}$ can be different for
different origin-destination  pairs $\rm {O}_{i}$--$\rm {D}_{j}$ of the network,
where $i=1,2,\ldots,I$ and $j=1,2,\ldots,J$). 
Before the network throughput maximization approach is applied to a large traffic or transportation network, 
  for each of the origin-destination  pairs $\rm {O}_{i}$--$\rm {D}_{j}$ of the network
	   a set of the alternative routes  should be found.
		
 \subsection{Network throughput maximization approach:
Prevention of breakdown in   networks at the maximum network throughput \label{Proc_BM_M}}

In   real traffic and transportation networks  
	the  network inflow rate $Q$ (\ref{OD_F})
	and, therefore, the flow rates $q^{(k)}_{\rm sum}$, $k=1,2,\ldots,N$ in the network increase 
	from  very small values (at night) to large values  during daytime. 
	At small initial values $Q$ and $q^{(k)}_{\rm sum}$    
	condition (\ref{C_min_q_in}) is valid for each of the network bottlenecks:
	\begin{equation} 
 q^{(k)}_{\rm sum} < C^{(k)}_{\rm min},  \quad   k=1,2,\ldots,N.
 \label{No_ind_all_BM}  
 \end{equation}
	In accordance with condition (\ref{C_min_q_in}), conditions
	(\ref{No_ind_all_BM}) mean that
	free flow  is stable with respect to traffic breakdown at each of the network
	bottlenecks. Therefore, no traffic breakdown can occur in the network.
 Due to the increase in $Q$,	the flow rate $q^{(k)}_{\rm sum}$
at least for one of the network
bottlenecks   
   $k=k^{(1)}_{1}$ becomes very close to $C^{(k)}_{\rm min}$. Therefore, in formula  
	\begin{equation} 
	 q^{(k)}_{\rm sum}+\epsilon^{(k)} 
  = C^{(k)}_{\rm min} \ {\rm for} \ k=k^{(1)}_{1}
	\label{Cap_min_1_BM} 
	\end{equation}
 a positive value 
$\epsilon^{(k)}=C^{(k)}_{\rm min}-q^{(k)}_{\rm sum}$
becomes very small:
	\begin{equation} 
	 \epsilon^{(k)}/C^{(k)}_{\rm min}\ll 1 \ {\rm for} \ k=k^{(1)}_{1}.
	\label{Cap_min_1_BM_epsilon}
	\end{equation}
	
		The approach to the   maximization  of the network throughput at 
which free flow conditions are ensured in the whole traffic or transportation network (network throughput maximization approach) is as follows.
We   maintain    condition (\ref{Cap_min_1_BM}) 
at the expense of the increase in   $q_{m}$
  on  other alternative routes.    
As a result,   condition
	 $q^{(k)}_{\rm sum}+\epsilon^{(k)} 
  = C^{(k)}_{\rm min}$
at $\epsilon^{(k)}/C^{(k)}_{\rm min}\ll 1$ is satisfied for another bottleneck 
$k=k^{(1)}_{2}$. 
When   $Q$ increases further, we repeat 
the above procedure  
for   other network bottlenecks. Consequently, we  get
			\begin{eqnarray} 
 \label{Cap_min_y_BM} 
 q^{(k)}_{\rm sum} +\epsilon^{(k)} = C^{(k)}_{\rm min} \ {\rm for} \
 k=k^{(1)}_{z} \
\\    z=1,2,\ldots,Z_{1}, \ {\rm where} \ Z_{1}\geq 1, \ Z_{1}\leq N, \nonumber   
     \end{eqnarray}
where		$\epsilon^{(k)}/C^{(k)}_{\rm min}\ll 1$.
Thus, when $Q$ increases, 
we   maintain  
conditions (\ref{Cap_min_y_BM})  at the expense of the increase
in   $q_{m}$ on other   
    alternative routes. 
		
		  However, 
	   due to the constrain $\lq\lq$alternative routes",  
  the number $Z_{1}$ of bottlenecks satisfying (\ref{Cap_min_y_BM}) is limited by
	some   value  $Z$ (where $Z\leq N$). All   values
$\epsilon^{(k)}$ in (\ref{Cap_min_y_BM}) are positive ones. Therefore,
conditions (\ref{Cap_min_y_BM}) are equivalent  to (\ref{No_ind_all_BM}).
Thus, due to the application of the network throughput maximization approach
 free flow  remains to be stable with respect to traffic breakdown
(F$\rightarrow$S transition) at each of the network
	bottlenecks. For this reason, no traffic breakdown can occur in the network.
	This means that conditions (\ref{Cap_min_y_BM}) in the   limit case   $Z_{1}=Z$  and
$\epsilon^{(k)}\rightarrow 0$  
	are related to {\it the  maximum possible
	network throughput} at which conditions (\ref{No_ind_all_BM})
	for the stability of free flow in the whole network are still satisfied.
	This is because 
at	the subsequent increase in the network inflow rate $Q$
the constrain $\lq\lq$alternative routes" does not allow us to maintain  
conditions (\ref{Cap_min_y_BM})  at the expense of the increase
in   $q_{m}$ on other   
    alternative routes: At least for one of the network bottlenecks
	conditions	(\ref{No_ind_all_BM})
cannot be satisfied any more. 

The latter means that when
    the  maximum  
	network throughput determined by conditions (\ref{Cap_min_y_BM}) in
	the   limit case   $Z_{1}=Z$  and
$\epsilon^{(k)}\rightarrow 0$ is exceeded, then
 at least for one of the network bottlenecks   condition 
 (\ref{C_min_q_in_larger}) is valid.
Under condition (\ref{C_min_q_in_larger}),	free flow is in a metastable state with respect
to traffic breakdown (F$\rightarrow$S transition) at a network bottleneck.
Therefore, traffic breakdown can occur in the network.

\subsection{A physical measure   of    traffic and transportation
		networks -- Network capacity \label{Proc_BM_C_min}}
	
	Within a steady-state analysis of
	traffic and transportation networks~\cite{Equil}, the limit case $Z_{1}=Z$
	in conditions (\ref{Cap_min_y_BM}) allows us
to define a network measure (or $\lq\lq$metric") -- {\it  network capacity}  
denoted by $C_{\rm net}$	as  follows. The   network capacity $C_{\rm net}$
	is the   maximum  total network inflow rate $Q$
	at which
	 conditions~\cite{Cal_C,Cal_C2} 
	\begin{eqnarray}
 \label{C_min_q_sum_equal} 
  q^{(k)}_{\rm sum} = C^{(k)}_{\rm min}  \  {\rm for}   	
\ k=k^{(1)}_{z},  	
\\  q^{(k)}_{\rm sum} < C^{(k)}_{\rm min} \ {\rm for} 
 \ k \ne k^{(1)}_{z},
 \nonumber	
	\\ 	  z=1,2,\ldots,Z; \ Z\geq 1, \  Z\leq N; \ k=1,2,\ldots,N \nonumber 
     \end{eqnarray}
	are satisfied.	
		From a comparison of conditions (\ref{Cap_min_y_BM}) and (\ref{C_min_q_sum_equal})
		we can see that
    the   total network inflow rate reaches the   network capacity~\cite{City}
		\begin{equation} 
 Q= C_{\rm net}
 \label{Cap_min_net_BM_equal}
 \end{equation}
when   the limit case
	$Z_{1}=Z$	in (\ref{Cap_min_y_BM}) is realized {\it and}
	all values $\epsilon^{(k)}$ in
	(\ref{Cap_min_y_BM})  are set to zero: $\epsilon^{(k)}=0$.
		In accordance with the definition of the network capacity (\ref{C_min_q_sum_equal})
	and condition
(\ref{C_min_q_in_larger}), at $Q= C_{\rm net}$ (\ref{Cap_min_net_BM_equal})
free flow becomes the metastable one with respect to traffic breakdown at     network bottleneck(s)
$k=k^{(1)}_{z}$ (\ref{C_min_q_sum_equal}).
This means that under conditions (\ref{C_min_q_sum_equal}) traffic breakdown can occur
in the network.

 		When  
\begin{equation} 
 Q< C_{\rm net},
 \label{Cap_min_smaller}
 \end{equation}
as explained above, conditions (\ref{Cap_min_y_BM}) 
are satisfied. Under conditions (\ref{Cap_min_y_BM}),
free flow is stable with respect to traffic breakdown (F$\rightarrow$S transition)
in the whole network. Therefore, no traffic breakdown can occur in the network.
  This explains the physical sense of the   network
capacity $C_{\rm net}$: 
Dynamic traffic assignment   in the network in accordance with the network throughput maximization approach
  maximizes the network throughput at which  traffic breakdown cannot occur   in the whole network.

Respectively, 	when   
\begin{equation} 
 Q> C_{\rm net},
 \label{Cap_min_net_BM_larger} 
 \end{equation}
 due to the constrain $\lq\lq$alternative routes" of the network throughput maximization approach
at least for one of the network bottlenecks the flow rate
 in free flow at the bottleneck becomes  larger than the minimum bottleneck
capacity:  $q^{(k)}_{\rm sum} > C^{(k)}_{\rm min}$. 
For this reason, rather than conditions (\ref{C_min_q_sum_equal})
  under condition (\ref{Cap_min_net_BM_larger})
	we get~\cite{Q_change}
	\begin{eqnarray}
 \label{C_min_q_sum_larger} 
  q^{(k)}_{\rm sum} > C^{(k)}_{\rm min} \ {\rm at} 
\ k=k^{(2)}_{w}, 
\\  q^{(k)}_{\rm sum}\leq C^{(k)}_{\rm min} \ {\rm at}    
 \ k \ne k^{(2)}_{w},
 \nonumber
		 \\  w=1,2,\ldots,W; \ W\geq 1, \  W\leq N; \ k=1,2,\ldots,N. \nonumber 
     \end{eqnarray}
		Therefore,  in accordance with
(\ref{C_min_q_in_larger}),
free flow is the metastable one with respect to traffic breakdown at the related network bottleneck(s).
This means that under conditions (\ref{C_min_q_sum_larger}) traffic breakdown can occur
in the network.

\subsection{The maximization of the  network throughput  
   in non-steady state of network \label{nonsteady_S}} 

 As above-mentioned, the definition of the 
network capacity (\ref{C_min_q_sum_equal}), (\ref{Cap_min_net_BM_equal}) is valid only within
 a steady-state analysis of
	the networks. Contrarily,
  conditions (\ref{Cap_min_y_BM}) at 
$Z_{1}=Z$ and  $\epsilon^{(k)}\rightarrow 0$, for which
free flow is stable with respect to traffic breakdown at each of the network bottlenecks,
are applicable even when free flow distribution in the network
cannot be considered a steady one. Indeed,
in conditions (\ref{Cap_min_y_BM})  
only local flow rates $q^{(k)}_{\rm sum}$ in free flow at network bottlenecks 
are used. 

Therefore,  conditions (\ref{Cap_min_y_BM})  
at which free flow conditions persist in the whole network
can be used for dynamic traffic assignment in real
 traffic and transportation networks under real non-steady state conditions, i.e.,
without involving an analysis of the   network capacity.

 We can see that the basic objective of the
 approach of the maximization of the
		network throughput   introduced in the paper
		is to guarantee that condition (\ref{C_min_q_in_larger}) is
  satisfied at {\it none} of the network bottlenecks.

	\section{Discussion \label{Dis_S}}
 
\subsection{The network throughput maximization approach
 versus Wardrop's equilibria
 \label{Wardrop_D_S}}

 From results of Ref.~\cite{BM_BM} in which a different application of the BM principle has been studied
(see Sec.~\ref{BM_versus_S}),
one can   assume  that the assignment procedure with the  network
 throughput maximization approach (\ref{Cap_min_1_BM})--(\ref{Cap_min_y_BM})
introduced in this article
 should give a better performance than assignment procedures designed through the use of Wardrop's UE or SO. 
However,   the following questions arise
that could not be answered in~\cite{BM_BM}:

(i) Even when the network throughput maximization approach (\ref{Cap_min_y_BM})
 exhibits the better performance in comparison with the Wardrop's equilibria, whether does 
 lead to {\it large enough benefits} that justify to use this approach instead 
of the Wardrop's equilibria? Indeed, it seems that  the use of the 
network throughput maximization approach (\ref{Cap_min_y_BM}) exhibits
 considerable disadvantages: Some of the drivers should use longer routes.
  
(ii) Is there a general measure for a comparison of 
 the performance of different assignment procedures for an arbitrary traffic network?

Both a general analysis and simulations
 of the network throughput maximization approach versus Wardrop's equilibria  are made below
 {\it only} for the case $Q<C_{\rm net}$
(\ref{Cap_min_smaller}), when the application of the  
network throughput maximization approach (\ref{Cap_min_y_BM})  ensures
 that traffic breakdown cannot occur in the network. One of the benefits
of this analysis is that we can find a general explanation of the
deterioration of 
traffic system through application of Wardrop's equilibria
that is independent on network characteristics.

\subsubsection{Deterioration of 
traffic system through application of Wardrop's equilibria: General analysis \label{BM_UE_S}}

		A real traffic or transportation
		network consists of alternative routes
	with very different lengths. Therefore, at small enough
	flow rates $q_{m}$, there are routes with short travel times ($\lq\lq$short routes")
	and routes with  longer travel times ($\lq\lq$long routes"). When $Q$ and, consequently, values 
	$q_{m}$ increase, the minimization of   travel times in the network   with the use of 
	dynamic traffic assignment based on
	Wardrop's UE or SO  leads to considerably larger increases in the flow rates
	on short routes in comparison with increases
	in the flow rates on long routes
	(e.g.,~\cite{Sheffi1984,BellIida19897,Peeta2001A,Mahmassani2001A,Rakha2009,DaganzoSheffi1977,Merchant1978A,Merchant1978B,Bell1992A,Bell2002A,Mahmassani1987A,Friesz2013A,Ben-Akiva2015A}).

Therefore, at  $Q< C_{\rm net}$
(\ref{Cap_min_smaller}), rather than conditions (\ref{Cap_min_y_BM}),
 any application of Wardrop's UE or SO  
leads to conditions $q^{(k)}_{\rm sum} < C^{(k)}_{\rm min}$ 
for some of the  bottlenecks  on    long routes, whereas for  
 some of the bottlenecks   on    short routes, 
we get $q^{(k)}_{\rm sum} > C^{(k)}_{\rm min}$. Therefore, in accordance with
condition (\ref{C_min_q_in_larger}), traffic breakdown can occur at network bottlenecks on the short routes.

One of the consequences of this general conclusion is
that already at  relatively small total network inflow rates  $Q< C_{\rm net}$
the application of the Wardrop's equilibria leads to the occurrence of   congestion in urban networks.
  An improvement of the performance of
the Wardrop's equilibria  with respect to the prevention of traffic breakdown in
a network could {\it not}
 be achieved, even if    
the	metastability of   free flow with respect to traffic breakdown at network bottlenecks 
 has been taken into account
	in travel time costs~\cite{Meta_Ward}. Indeed,
due to
the metastability that is realized under condition (\ref{C_min_q_in_larger}), 
   travel time costs can exhibit
  discontinuities when traffic breakdown has occurred. However,
 possible discontinuities in travel time costs
cannot change the above conclusion that the application of the Wardrop's
equilibria at  $Q\rightarrow C_{\rm net}$
  does result in the metastability of free flow
at some of the network bottlenecks: Under free flow conditions, in accordance with the
Wardrop's  UE or SO  the flow rate on a short route
	should be larger than the flow rate on a long route. This is
	independent on  
  values $C^{(k)}_{\rm min}$ for bottlenecks on the route.
Contrarily,
 in accordance with conditions (\ref{Cap_min_y_BM}) resulting 
from the network throughput maximization approach,
the flow rate at any network bottleneck $q^{(k)}_{\rm sum}$ 
should be smaller than the minimum capacity
$C^{(k)}_{\rm min}$ of the bottleneck. This  does not
depend  on 
whether a bottleneck is on a short  route    or it is on a long route.
 
 The above conclusion that at $Q\rightarrow C_{\rm net}$ 
		any application of the Wardrop's equilibria results in the occurrence of the metastable free flow
		at some of the network bottlenecks can additionally be explained through a consideration of
  the following
		hypothetical network: In the network with  different   lengths of alternative routes,
	values of minimum capacity $C_{\rm min}$ are assumed the same for
	any bottleneck. We assume that our statement about the metastability of free flow at some of the network
	bottlenecks
  might be incorrect:
	An
	application of Wardrop's UE or SO might result in conditions (\ref{Cap_min_y_BM}) 
at which free flow is stable with respect to traffic breakdown at each of the network bottlenecks.
 However, 
at   $C^{(k)}_{\rm min}=C_{\rm min}$ 
and $\epsilon^{(k)}\rightarrow 0$ in  (\ref{Cap_min_y_BM}),
on   routes with 
  different lengths the flow rates $q^{(k)}_{\rm sum}$
	 must be the same
  for all  bottlenecks for which conditions (\ref{Cap_min_y_BM})
	are satisfied. 
	This contradicts   the sense of any application
	of Wardrop's  UE or SO: On average, the flow rates
	  should be larger on short routes then those on
	long routes.

 Thus, due to the application of Wardrop's UE or SO,
already at  
$Q< C_{\rm net}$  (\ref{Cap_min_smaller}) it 
turns out that
for some of the network bottlenecks
$q^{(k)}_{\rm sum} > C^{(k)}_{\rm min}$. Therefore, traffic breakdown can occur  
at   these   bottlenecks.
In this case, it is not possible to predict the time instant at which
traffic breakdown occurs at these  bottlenecks. This is
because the time delay to the breakdown
$T^{\rm (B)}$ is a random value~\cite{Kerner_Review,KernerBook}.
We can apply the Wardrop's UE or SO for dynamic traffic assignment
	without any delay after the breakdown has occurred. This is possible because
	the speed decrease due to the breakdown can be measured.
  However, after  the assignment has been made,
	there is always a time delay in the change
of traffic variables  at the bottleneck location. This time delay is
 caused by travel time from the beginning of a link to
 a bottleneck location on the link.  
  Therefore,    
it is not possible to 
  avoid  congested traffic occurring due to the breakdown.
	A control method can only 
effect on a spatiotemporal distribution of congestion in the network. 

From this general analysis, we can see that the main benefit of the network throughput maximization approach
in comparison with the Wardrop's equilibria is as follows:
Even when the total network inflow rate $Q$ approaches the   network capacity, i.e.,
at $Q\rightarrow C_{\rm net}$
	the network throughput maximization approach
		does ensure free flow conditions at which traffic breakdown cannot occur
		in the whole network. Contrarily, at $Q\rightarrow C_{\rm net}$
	  any application of the Wardrop's equilibria
		does lead to congested traffic in the network.

The physics of traffic breakdown in traffic and transportation networks
revealed in Secs.~\ref{Proc_BM_M}  and~\ref{Proc_BM_C_min}  
shows   that the wish of humans to use shortest   routes of a network
contradicts fundamentally
 the physical nature of traffic breakdown in the network.
Therefore, in contrast to the human wish, 
the use of the classical Wardrop's equilibria 
results basically in the occurrence of congestion in urban networks. 
 This can explain
why   approaches to dynamic traffic assignment based on the 
classical  Wardrop's UE or SO equilibria (e.g.,~\cite{Sheffi1984,BellIida19897,Peeta2001A,Mahmassani2001A,Rakha2009,DaganzoSheffi1977,Merchant1978A,Merchant1978B,Bell1992A,Bell2002A,Mahmassani1987A,Friesz2013A,Ben-Akiva2015A}) have failed by their applications in the real world.
		
		\subsubsection{Numerical simulations \label{Numerical_S}}

		\begin{figure}
 \begin{center}
 \end{center}
\includegraphics[width=8 cm]{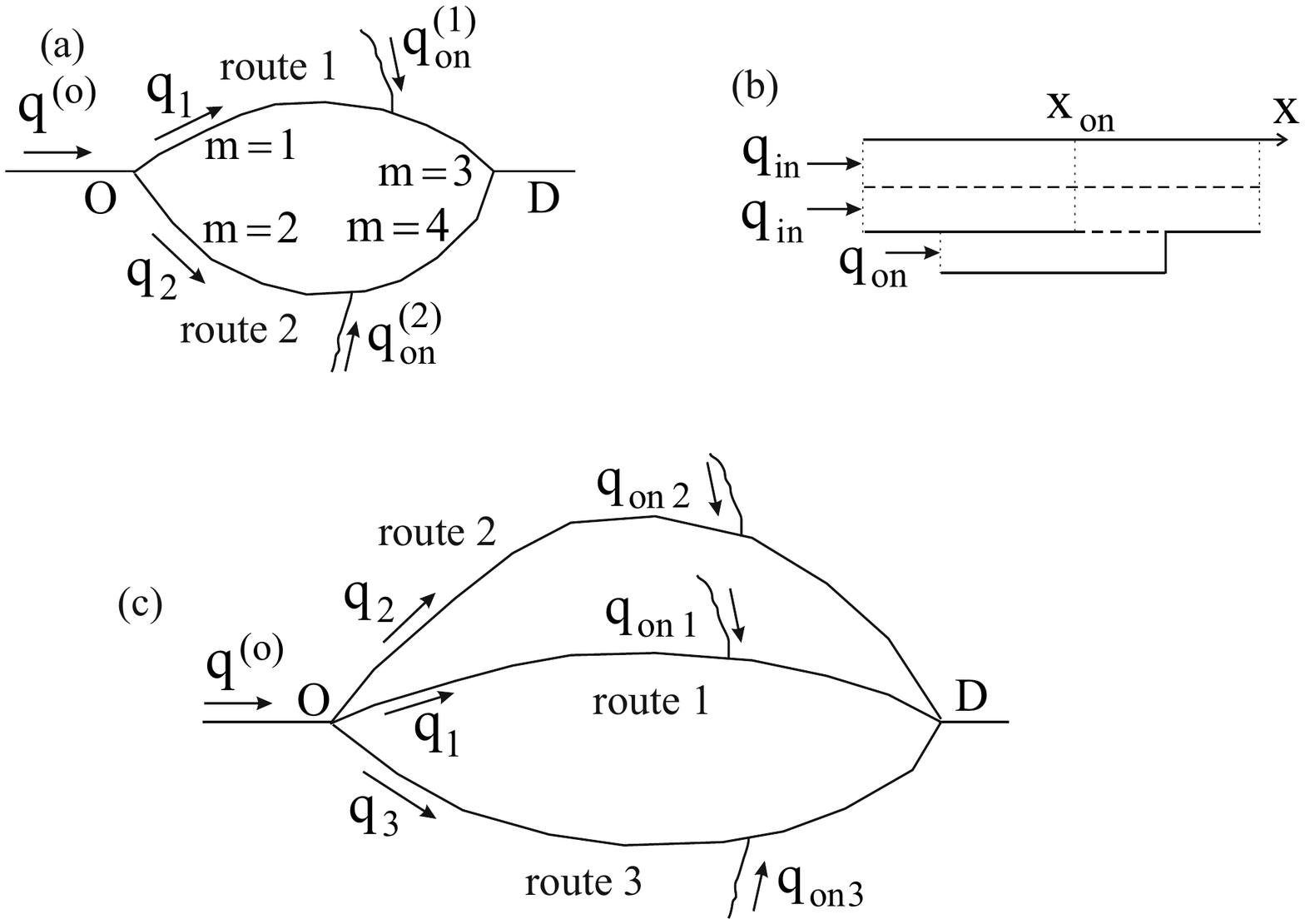}
\caption{Model of two-route and three-route networks:
(a, b) Sketch of a simple network with two routes and two on-ramp bottlenecks with 
$L_{1}=$ 20 km, $L_{2}=$ 25 km   (a) and model of on-ramp bottleneck on 
a two-lane road (b);
the beginning of the on-ramp merging region is at $x_{\rm on}=$ 15 km (b);
$q^{(1)}_{\rm on}=$ 400 vehicles/h, $q^{(2)}_{\rm on}=$ 700 vehicles/h; 
calculated values $C^{\rm (k)}_{\rm min}=$ 3980 for $k=1$, 3760  vehicles/h for $k=2$.
(c) Sketch of  network with three routes and three on-ramp bottlenecks that model is shown in (b);
$L_{1}=$ 20 km,  $L_{2}=$ 25 km,  $L_{3}=$ 22.5 km;
given on-ramp inflow rates
$q^{(k)}_{on}=$
  300 vehicles/h for route 1 ($k=1$),
 800 vehicles/h for route 2 ($k=2$), and 
 500 vehicles/h for route 3 ($k=3$);
calculated values $C^{\rm (k)}_{\rm min}=$ 4000 for $k=1$, 3740 for $k=2$,
and 3850  vehicles/h for $k=3$.  
\label{BM_Network_Model2_new} } 
\end{figure}

       \begin{figure}
\begin{center}
\includegraphics[width=8 cm]{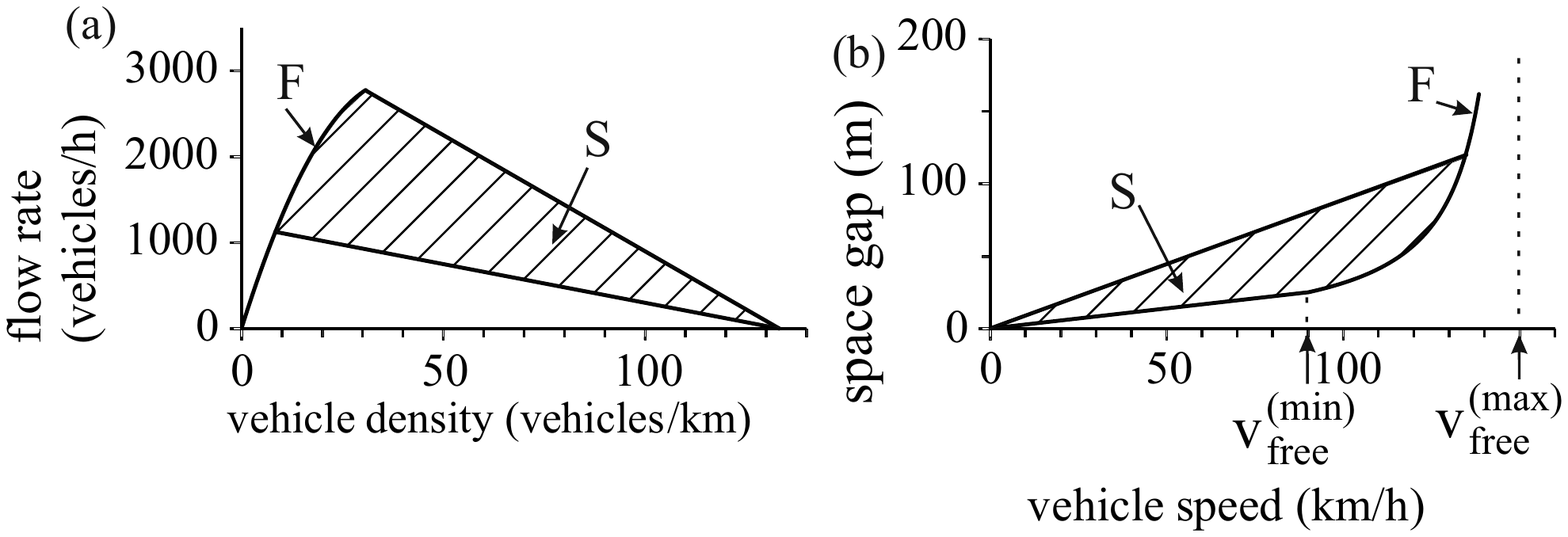}
\caption{Model steady states in Kerner-Klenov model  
  in the flow-density plane (a) and in the space-gap-speed plane (b);
F -- free flow, S -- synchronized flow; maximum free flow speed
is given by formula (\ref{free_speed_BM}).
  \label{KKl_2D_regions} } 
\end{center}
\end{figure}

  The Wardrop's UE or SO  and
   the network throughput maximization approach are devoted to  the optimization of large, complex vehicular traffic networks. 
However, to illustrate the general conclusions made above, it is sufficient to
 simulate 
simple models of   traffic networks, two-route network with two on-ramp bottlenecks (Fig.~\ref{BM_Network_Model2_new} (a, b))~\cite{TwoRoute}  
  and three-route network 
	with with three on-ramp bottlenecks
	(Fig.~\ref{BM_Network_Model2_new} (c))~\cite{Num_Sim}.
 In both networks,
 vehicles move from origin $O$ to destination $D$
 (Figs.~\ref{BM_Network_Model2_new} (a, c)). 
 Each of the network routes   is a two-lane road with 
  an on-ramp bottleneck (Figs.~\ref{BM_Network_Model2_new} (a--c)).
	In two-route network (Fig.~\ref{BM_Network_Model2_new} (a)), there are two  
alternative routes  1 and 2 with lengths
 $L_{1}$ and $L_{2}$ (with $L_{2}>L_{1}$).
In three-route network (Fig.~\ref{BM_Network_Model2_new} (c)), there are    three  
alternative routes  1, 2, and 3 with lengths
 $L_{1}$, $L_{2}$, and $L_{3}$ (with $L_{2}>L_{3}>L_{1}$).
	The total network inflow rate is equal to
$Q= q^{\rm (o)} +\sum^{N}_{k=1} q^{(k)}_{on}$,   
where $N=$ 2 and 3, respectively,
for two-route and three-route networks (Figs.~\ref{BM_Network_Model2_new} (a, c)).
 In free flow,	due to  a complex dynamics of
 permanent speed disturbances  at 
the   bottlenecks  as well as a 
decreasing-dependence of the vehicle speed   on the vehicle density,   route travel times   
depend considerably
on the link inflow rates  $q_{m}$  ($m=1, 2$ for Fig.~\ref{BM_Network_Model2_new} (a)
and  $m=1, 2, 3$ for Fig.~\ref{BM_Network_Model2_new} (c)).  
	The   flow rates downstream of the bottlenecks are
	 $q^{(k)}_{sum}=q_{k}+ q^{(k)}_{on}$, 
where on-ramp inflow rates $q^{(k)}_{on}$ are given constants
(see caption of Fig.~\ref{BM_Network_Model2_new}),
 $q_{k}=q_{m}$ and
$k=m$
 ($m=1, 2$ in Fig.~\ref{BM_Network_Model2_new} (a, c)).
	
	For simulations, we use  
 the Kerner-Klenov stochastic microscopic three-phase traffic flow 
model~\cite{KKl,KKl2009A}, 
 in which the maximum speed of vehicles in free flow is a function
of the space gap  $g$  between vehicles  (Fig.~\ref{KKl_2D_regions}).
Dependence of the maximum
free flow speed on the space gap $g$ between vehicles 
 is given by formula:
\begin{equation}
v_{\rm free}(g)= \max(v^{(\rm min)}_{\rm free}, \ v^{(\rm max)}_{\rm free}(1-k_{\rm v}d/(d+g))),
\label{free_speed_BM}
\end{equation}
where $v^{\rm (min)}_{\rm free}=90$ km/h, $v^{\rm (max)}_{\rm free}=150$ km/h, $d=7.5$ m (vehicle length),
and $k_{\rm v}=1.73$.
The model incorporates the metastability of free flow  with respect to an F$\rightarrow$S transition
(traffic breakdown) at network bottlenecks~\cite{KernerBook,KernerBook2,KKl,KKl2009A}.
 Simulations show that
  this stochastic model allows us to take into account the metastability in
	travel time costs. With the use of simulations, this model feature   
	 illustrates a general conclusion
	made in Sec.~\ref{BM_UE_S} that the incorporation of the metastability in
	travel time costs does not prevent the occurrence of congestion
	by the use of the Wardrop's equilibria
	at  
$Q< C_{\rm net}$  (\ref{Cap_min_smaller}).
Because the physics of 
this model has already been studied in details~\cite{KernerBook,KernerBook2,KKl,KKl2009A},
the model is given in Appendix~\ref{App}.

	Simulations show that for these network models
	the  network capacity
	\begin{equation}
C_{\rm net}=\sum^{N}_{k=1} C^{\rm (k)}_{\rm min}.
\label{minC}
\end{equation}
At chosen network parameters
(see captions to Fig.~\ref{BM_Network_Model2_new}),
  $C_{\rm net}=$7740 vehicles/h for two-route and
 to $C_{\rm net}=$11590 vehicle/h for three-route networks, respectively.
As long as $Q< C_{\rm net}$,
when the network throughput maximization approach   is applied, then   
no traffic breakdown
 occurs in both networks.

 \begin{figure}
\begin{center}
\includegraphics[width=8 cm]{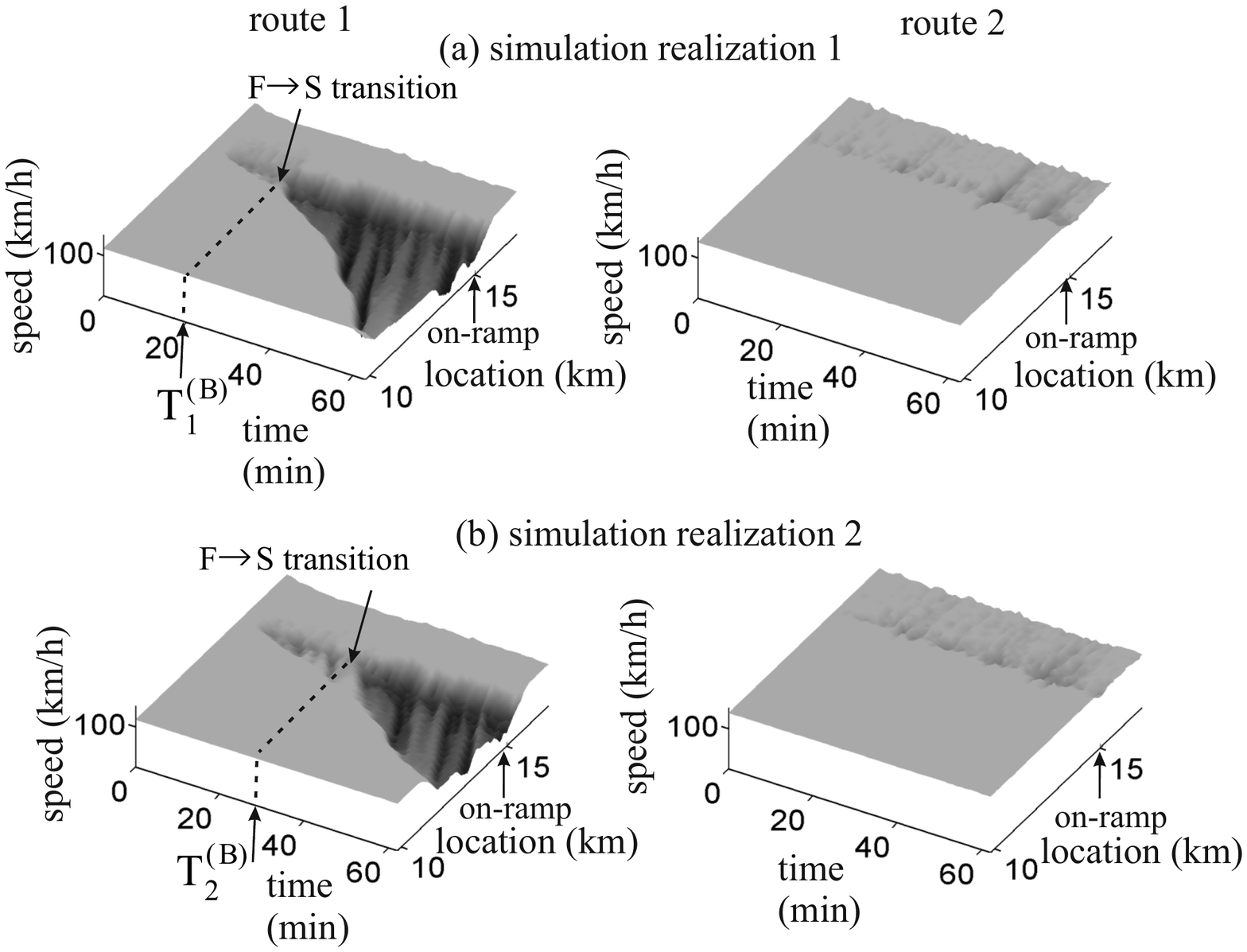}
\caption{Simulations of static traffic assignment with the Wardrop's  UE  
for  network  in Fig.~\ref{BM_Network_Model2_new} (a). 
(a--d) Speed in space and time on routes 1 (left panel) and 2 (right panel)
for two simulation realizations  with time delays of the breakdown
$T^{\rm (B)}=$ 20 min (realization 1)
and $T^{\rm (B)}=$ 29 min (realization 2).  $Q= 7000$ vehicles/h. 
In accordance with (\ref{UE_eq2}),
 $q_{1}=4060$ vehicles/h, $q_{2}=1840$ vehicles/h.
Arrows F$\rightarrow$S  show time instants of F$\rightarrow$S 
transitions.
\label{UE_Prob_5900_short} }  
 \end{center}
\end{figure}

	 \begin{figure}
\begin{center}
\includegraphics[width=8 cm]{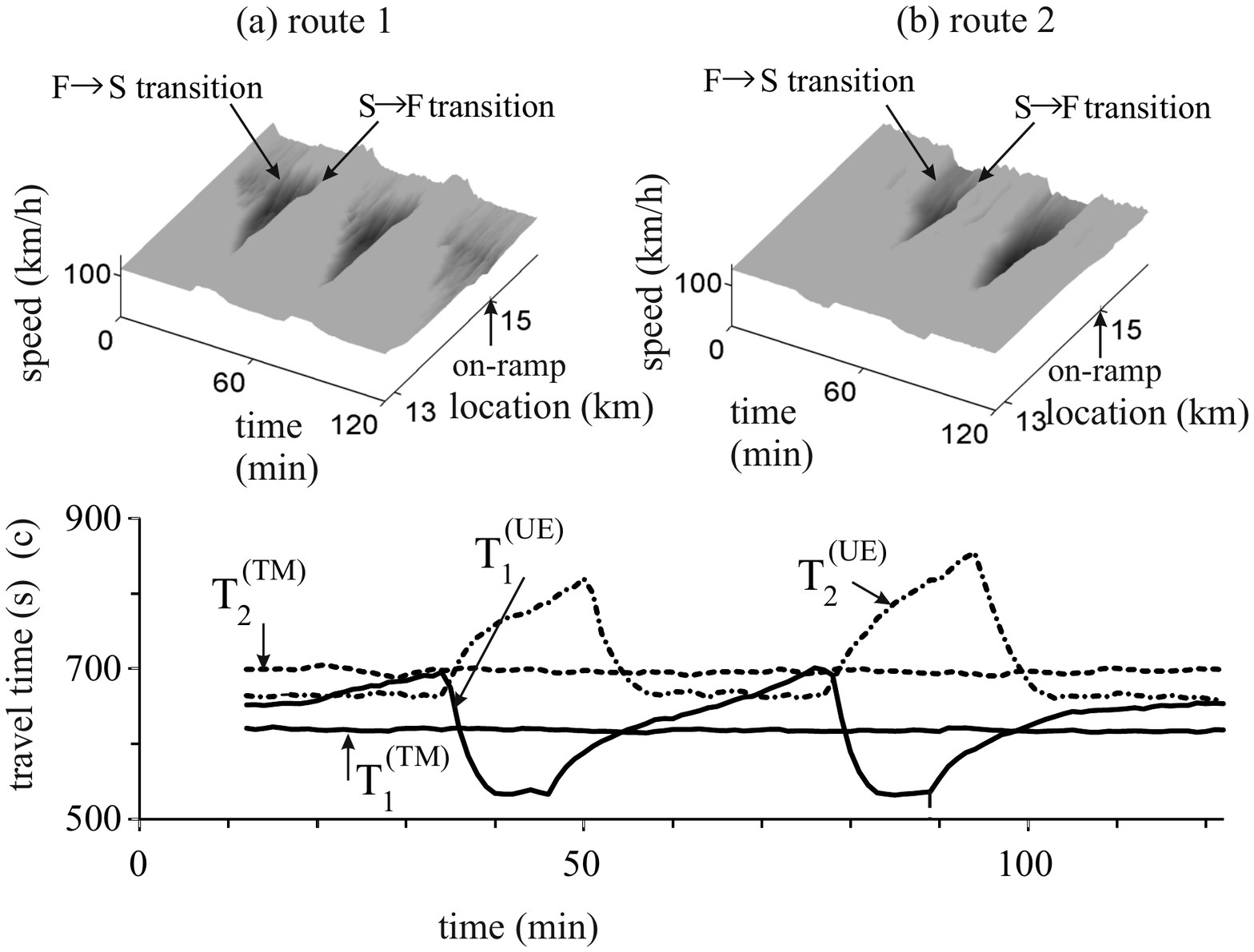}
\caption{Simulations of dynamic traffic assignment
for two-route network  (Fig.~\ref{BM_Network_Model2_new} (a))
at time-independent inflow rate $q^{\rm (o)}$. 
(a, b) Speed in space and time on routes 1 (a) and 2 (b) under application of Wardrop's UE.
(c) Travel times  
 $T_{1}=T^{\rm (UE)}_{1}(t)$, $T_{2}=T^{\rm (UE)}_{2}(t)$ for the Wardrop's  UE principle,
$\Delta q=$ 2000    vehicles/h~\cite{ANCONA}. Travel times
  $T_{1}=T^{\rm (TM)}_{1}$, $T_{2}=T^{\rm (TM)}_{2}$ are related to  
application of the network throughput maximization approach for which
 $q_{1}=3560$ vehicles/h, $q_{2}=2340$ vehicles/h.
    $Q=$ 7000 vehicles/h.
\label{UE_ANCONA_5900_short} } 
 \end{center}
\end{figure}

\begin{figure}
\begin{center}
\includegraphics[width=8 cm]{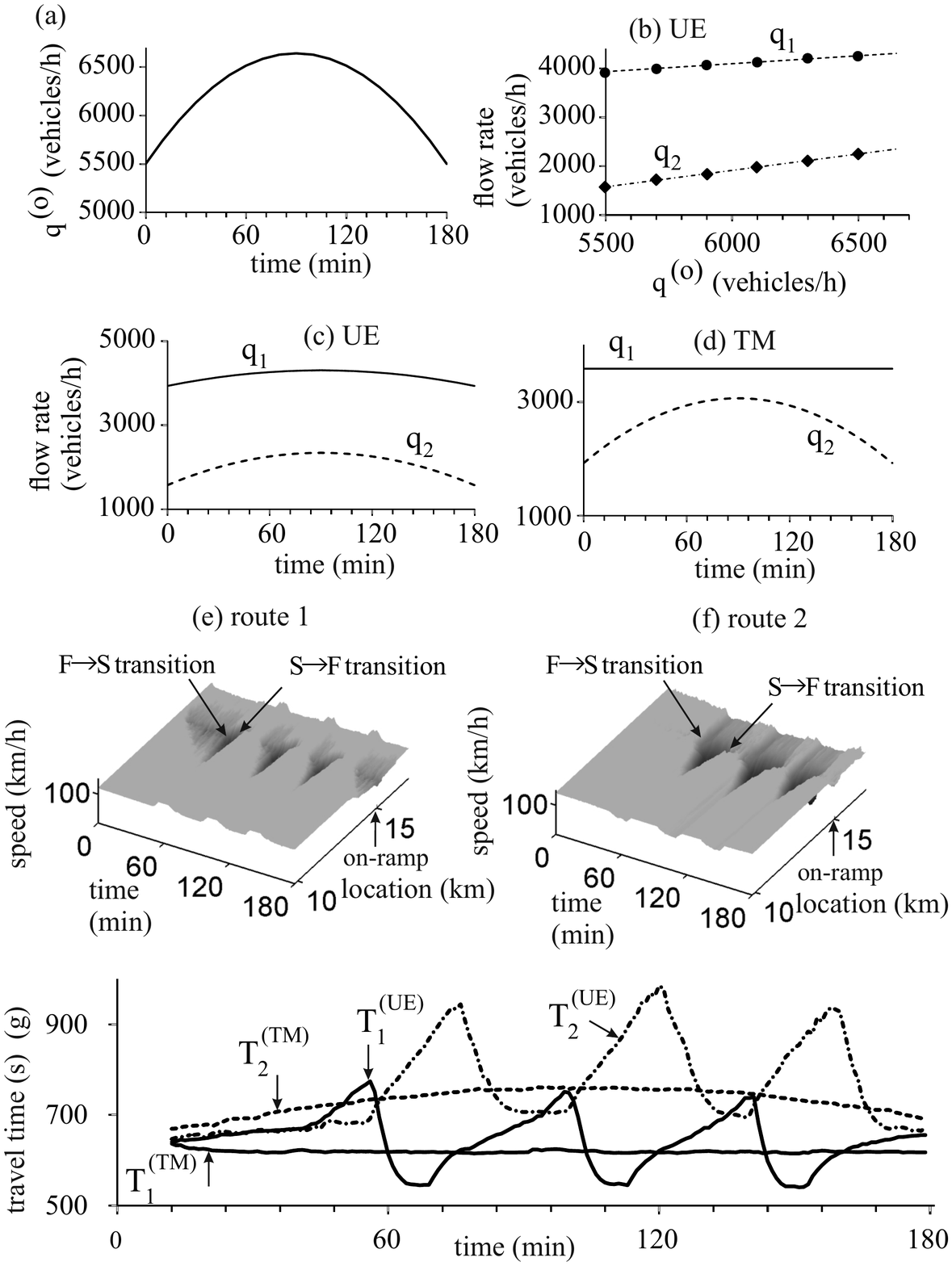}
\caption{Simulations of dynamic traffic assignment
for two-route network  (Fig.~\ref{BM_Network_Model2_new} (a))
at time-dependent inflow rate $q^{\rm (o)}(t)$ (a);
the maximum value of this flow rate $q^{\rm (o)}_{\rm max}=6636$ vehicles/h
satisfies condition (\ref{Cap_min_smaller}).
(b, c) Flow rates $q_{m}$,  $m=1,2$ for Wardrop's UE as functions of $q^{\rm (o)}$ (b)
and on time (c)
calculated  
 in accordance with  (\ref{UE_eq2})~\cite{UE_T}; in (b),
calculated points are well fitted by lines $q_{1}=0.33q^{\rm (o)}+2170$
  and $q_{2}=0.67q^{\rm (o)}-2170$ vehicles/h.
 (d) Time-dependent flow
rates $q_{m}(t)$  $m=1,2$
 for the network throughput maximization approach.
(e, f) Speed in space and time on routes 1 (e) and 2 (f)
under application of Wardrop's UE.
(g) $T_{r}=T^{\rm (UE)}_{r}(t)$ on different routes $r=1,2$  for the Wardrop's  UE principle,
$\Delta q=$ 2000    vehicles/h~\cite{ANCONA}. In (g), travel times
  $T_{r}=T^{\rm (TM)}_{r}(t)$ on different routes $r=1,2$ are related to  
application of the network throughput maximization approach at which no traffic breakdown
occurs at the bottlenecks.
\label{UE_ANCONA_2} } 
 \end{center}
\end{figure}

\begin{figure}
\begin{center}
\includegraphics[width=8 cm]{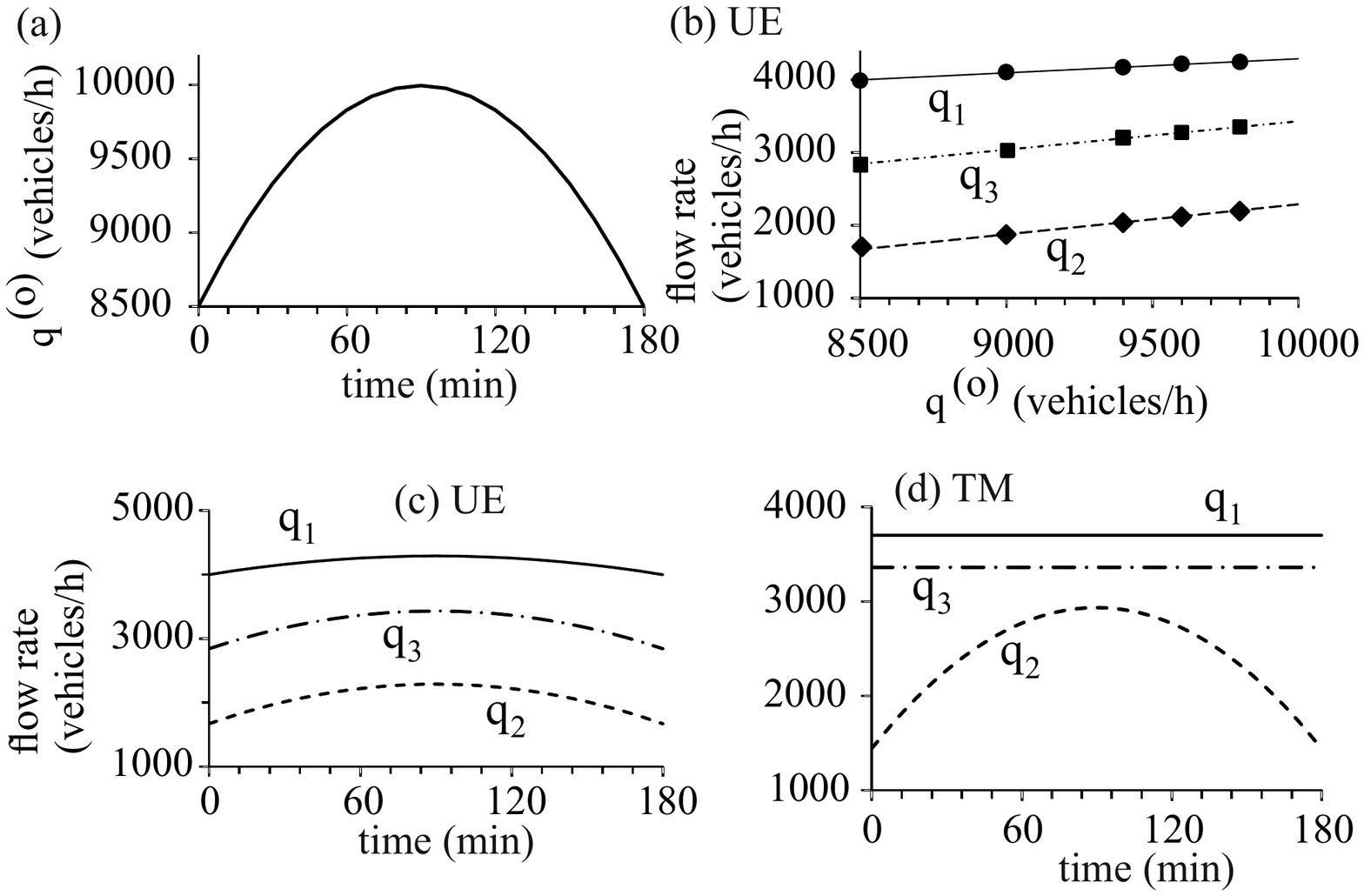}
\caption{Simulations of distribution of flow rates between different routes
through dynamic traffic assignment
in three-route network  (Fig.~\ref{BM_Network_Model2_new} (c))
at time-dependent inflow rate $q^{\rm (o)}(t)$ (a);
the maximum value of this flow rate $q^{\rm (o)}_{\rm max}=$ 9994 vehicles/h
satisfies condition (\ref{Cap_min_smaller}). 
(b, c) Flow rates $q_{m}$,  $m=1,2,3$ for Wardrop's UE as functions of $q^{\rm (o)}$ (b)
and on time (c)
calculated  
 in accordance with  (\ref{UE_eq3})~\cite{UE_T}; in (b),
calculated points are well fitted by lines $q_{1}=0.19q^{\rm (o)}+2345$,
  $q_{3}=0.41q^{\rm (o)}-1850$,
  and   $q_{3}=0.39q^{\rm (o)}-495$ vehicles/h. (d) Time-dependent flow
rates $q_{m}(t)$  $m=1,2,3$
 for the network throughput maximization approach.
\label{UE_ANCONA_3} } 
 \end{center}
\end{figure}

\begin{figure}
\begin{center}
\includegraphics[width=8 cm]{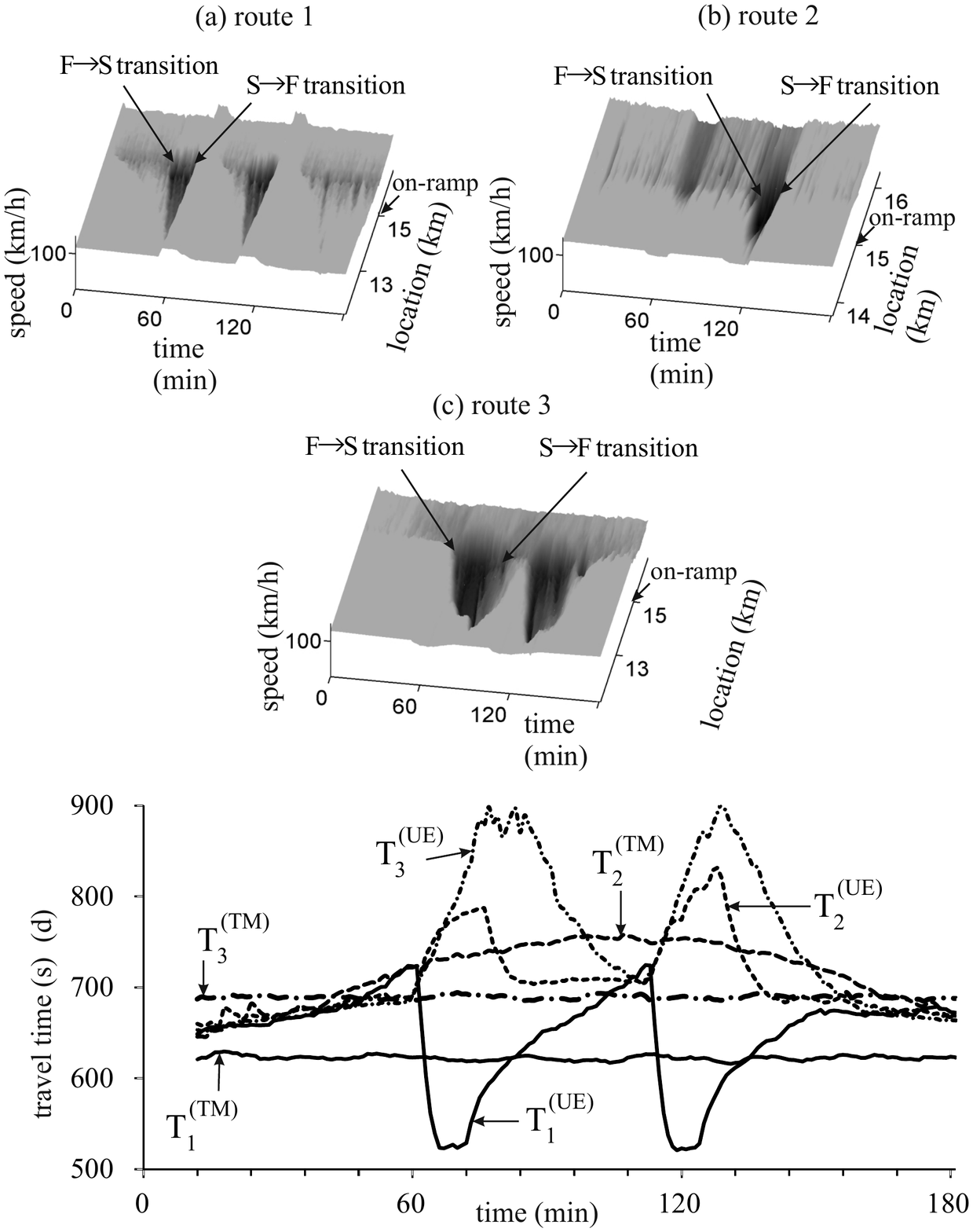}
\caption{Simulations of dynamic traffic assignment in three-route network 
accordingly to the 
  flow rates on different routes  that calculations are 
		shown in Fig.~\ref{UE_ANCONA_3}:
(a--c) Speed in space and time on routes 1 (a), 2 (b), and 3 (c)
under application of Wardrop's UE.
(d) $T_{r}=T^{\rm (UE)}_{r}(t)$ on different routes ($r=1,2,3$)  for the Wardrop's  UE principle,
$\Delta q=$ 2400    vehicles/h~\cite{ANCONA}. In (d), travel times
  $T_{r}=T^{\rm (TM)}_{r}(t)$ ($r=1,2,3$)  are related to  
application of the network throughput maximization approach at which no traffic breakdown occurs at the bottlenecks.
\label{UE_ANCONA_3_2} } 
 \end{center}
\end{figure}

	The Wardrop's UE for the two-route network shown in Fig.~\ref{BM_Network_Model2_new} (a) results in
\begin{equation}
T_{1}(q_{1}, q^{(1)}_{\rm on})=T_{2}(q_{2}, q^{(2)}_{\rm on}), \  q^{\rm (o)}=\sum^{2}_{m=1} q_{m},
\label{UE_eq2}
\end{equation} 
where $T_{r}, \ r=1,2$   are travel times at route $r=$1 and $r=$2, respectively.
For the three-route network (Fig.~\ref{BM_Network_Model2_new} (c)) under conditions
$q_{m}>0, \ m=1,2,3$ that are realized in simulations, we get
	\begin{equation}
T_{1}(q_{1}, q^{(1)}_{\rm on})=T_{2}(q_{2}, q^{(2)}_{\rm on})=T_{3}(q_{3}, q^{(3)}_{\rm on}),
 \  q^{\rm (o)}=\sum^{3}_{m=1} q_{m},
\label{UE_eq3}
\end{equation}
where $T_{r}, \ r=1,2,3$   are travel times at route 1,2, and 3, respectively.
	To disclose the physics of the dynamic traffic assignment with Wardrop's UE under condition
(\ref{Cap_min_smaller}), we consider and compare
both a hypothetical case of
a time-independent  
 inflow rate $q^{\rm (o)}=\sum^{M}_{m=1} q_{m}$
at the origin  of the network  (Figs.~\ref{UE_Prob_5900_short}
and~\ref{UE_ANCONA_5900_short}) with a more realistic case of a time-dependent inflow rate
 $q^{\rm (o)}(t)$
(Figs.~\ref{UE_ANCONA_2}--\ref{UE_ANCONA_3_2}). In Figs.~\ref{UE_ANCONA_2}
and~\ref{UE_ANCONA_3}, in accordance with almost all empirical observations
we simulate a morning (or evening) rush hour 
 in which the inflow rate $q^{\rm (o)}(t)$
 firstly increases and then decreases over daytime (Figs.~\ref{UE_ANCONA_2} (a)
and~\ref{UE_ANCONA_3} (a))~\cite{UE_T}.

 For two-route network (Fig.~\ref{BM_Network_Model2_new} (a)), route 1 is shorter than route 2. 
To satisfy   (\ref{UE_eq2}), the flow rate $q_{1}$  
should be larger than the flow rate $q_{2}$. For this reason,
 already at $Q=7000<C^{\rm (net)}_{\rm min}=7740$ vehicles/h, the probability of traffic 
breakdown at bottleneck 1 is equal to $0.59$.
Therefore, under application of the {\it static} dynamic
assignment with Wardrop's UE (flow rates   $q_{m}$ that
 satisfy (\ref{UE_eq2}) at $t=0$
 do not depend on time)
 we have found a random time-delayed traffic 
breakdown (F$\rightarrow$S
transition) at bottleneck 1 leading to traffic congestion (Fig.~\ref{UE_Prob_5900_short}).

	The   application of the Wardrop's UE  for
	{\it dynamic} traffic assignment~\cite{ANCONA}  results in a   {\it random 
	process} of the congested pattern emergence 
	due to an F$\rightarrow$S transition with the subsequent dissolution of the   pattern
	due to a return S$\rightarrow$F transition, 
	and so on (Fig.~\ref{UE_ANCONA_5900_short}):  In different simulation realizations,
	we have found different sequences of the congested pattern emergence and dissolution. 
	As in many other applications of the Wardrop's UE~\cite{Wahle,Davis1,Davis2},
	this random process leads to large oscillations of travel times
	$T^{\rm (UE)}_{1}$ and $T^{\rm (UE)}_{2}$ 
	(Fig.~\ref{UE_ANCONA_5900_short} (c)), whereas travel times
	$T^{\rm (TM)}_{1}$ and $T^{\rm (TM)}_{2}$ under the use of the network throughput maximization approach are 
	  time-independent~\cite{SO}. 
		
		In  Fig.~\ref{UE_ANCONA_2},
		the  total network inflow rate $Q(t)$ depends on time and it does not exceed
		the   network capacity. We can see that even in this case under application of the Wardrop's UE
		we get qualitatively the same random   process of sequences
		of the congested pattern emergence and dissolution with
		large oscillations of travel times. 
		 Due to the application of the network throughput maximization approach,   condition
		(\ref{Cap_min_1_BM})   is satisfied for bottleneck 1; therefore, the increase in $q^{\rm (o)}(t)$ 
		leads to the increase in $q_{2}$ on the alternative route 2 only.
		For this reason, from Fig.~\ref{UE_ANCONA_2} (d, g)  we can see that as long as
		the inflow rate $q^{\rm (o)}$ increases over time
		the flow rate $q_{1}$ and travel time
		$T^{\rm (TM)}_{1}$
		does not depend on time, whereas $q_{2}$ and $T^{\rm (TM)}_{2}$ increase over time.
		
		In  Fig.~\ref{UE_ANCONA_3} (a),
		the  total network inflow rate $Q(t)$ depends also on time and it does not exceed
		the   network capacity for three-route network (Fig.~\ref{BM_Network_Model2_new} (c)). 
		A  more complex
		network structure results in a more sophisticated  spatiotemporal distribution of congestion between
		different routes (Fig.~\ref{UE_ANCONA_3_2} (a--c)). Nevertheless,
		we find   qualitatively  the same random   process of sequences
		of the congested pattern emergence and dissolution on different routes with
	 oscillations of route travel times (Fig.~\ref{UE_ANCONA_3_2} (d)) as that 
		in a simpler two-route network (Fig.~\ref{UE_ANCONA_2}).
		A difference with two-route network is also found by the application of the network throughput maximization approach:
		For the flow rate $q^{\rm (o)}$ 
		used in Fig.~\ref{UE_ANCONA_3} (a) for three-route network,  
    conditions
		(\ref{Cap_min_y_BM})  are   satisfied for bottlenecks
		1 and 3. Therefore, the whole increase in $q^{\rm (o)}(t)$
		leads to the increase in $q_{2}$ on the alternative route 2 only (Fig.~\ref{UE_ANCONA_3} (d)).

\subsubsection{About application of network throughput maximization approach
  for real traffic and transportation networks \label{Ben_S}}

When the network throughput maximization approach is applied,
route travel times determine only the constrain $\lq\lq$alternative routes" that
		prevents the use of too long routes. After these
		long routes have been
		excluded   from dynamic traffic assignment with the network throughput maximization approach,
		the assignment is  determined by  conditions (\ref{Cap_min_y_BM}) 
		as explained in Sec.~\ref{Proc_BM_M}.
     Thus it seems that
		the use of the network throughput maximization approach in real traffic and transportation networks 
		has considerable disadvantages in comparison with applications of the Wardrop's equilibria: Some of the drives should use longer routes to avoid congestion in the network.
		
		However, possible benefits of the application of the network throughput maximization approach
		for the future organization of traffic and transportation networks
		can overcome
		these disadvantages. Indeed,   empirical analysis of real field microscopic traffic data shows that
		fuel consumption and, therefore, CO$_{2}$ emission 
		can be about 3--3.5 times larger in congested traffic than those in free flow
		(e.g.,~\cite{Hemmerle2015} and references there).
  Thus the
		maintenance of free flow conditions
		in urban areas  of industrial countries through the application of the network throughput maximization approach
		as explained in this article can contribute to the   environment protection
		against the  increase in CO$_{2}$ emission in the world.
		
		Moreover, through the use of short routes in applications of 
		the Wardrop's equilibria  congested traffic in a  network occurs.
		Travel times can  considerably increase due to congestion on short routes
		(Figs.~\ref{UE_ANCONA_5900_short},
		\ref{UE_ANCONA_2}, and~\ref{UE_ANCONA_3_2}). Therefore, the use of short routes
		does not necessarily lead to the reduction of travel time costs
		in real networks.
		
To maintain
	free flow conditions in urban networks for such an 
environment protection, already now there are  technical possibilities~\cite{Kerner_CO2}.
Through communication of  GPS vehicle data to a traffic control
 center, the center can provide
  appropriate information to the drivers as well as  organize
	an efficient network organization with 
	the network throughput maximization approach as introduced Secs.~\ref{Proc_BM_M} and~\ref{nonsteady_S}:  
	  (i) Traffic center can   store 
		characteristics of traffic breakdown at network bottlenecks found from
		measurements of traffic variables 
			with road detectors, video cameras and/or GPS probe vehicles (FCD -- floating car data).
(ii) Traffic   center can inform drivers individually about an eco-route
		calculated with the approach of Secs.~\ref{Proc_BM_M} and~\ref{nonsteady_S}.
		(iii) Electronic road charge systems based on GPS vehicle data   can facilitate
		the use of alternative routes associated with the network throughput maximization approach   for 
		the maintenance of conditions 
		 (\ref{Cap_min_y_BM}).    
		For example, as very small network inflow rates $Q$, when condition (\ref{Cap_min_1_BM}) is not satisfied,
		vehicles can freely choose routes without any (or the same) road charge.
		At larger values of $Q$, when conditions (\ref{Cap_min_y_BM})
		are satisfied for some of the network bottlenecks, 
		the vehicles using alternative (longer) routes should pay considerably smaller road charge
		(or no charge at all) than those using the routes 
		via network bottlenecks for which conditions (\ref{Cap_min_y_BM})
		are satisfied.
		
		\subsection{Network throughput maximization approach 
		as an application of breakdown minimization (BM) principle \label{BM_versus_S}}
		
		In~\cite{BM_BM,BM_BM2},   a breakdown minimization (BM) principle
	for dynamic traffic assignment and control
		in traffic and transportation networks	has been introduced.
		 The BM principle  states that the	 optimum of a traffic or transportation network with $N$
    bottlenecks is reached,
  when dynamic traffic assignment, optimization and/or  control are performed in the network 
  in such a way that the probability $P_{\rm net}$ for the  occurrence of   
	traffic breakdown 
  in at least one of the network bottlenecks during a given observation time
	interval $T_{\rm ob}$ reaches
   the minimum possible value.  The BM principle reads as follows~\cite{BM_BM,BM_BM2}:
	\begin{eqnarray}
\min \limits_{q_{1},q_{2},...,q_{\rm M},{\bf R}_{1},{\bf R}_{2},...,{\bf R}_{\rm N},  
\alpha_{1},\alpha_{2},...,\alpha_{N}} \{P_{\rm net}(q_{1},q_{2},...,q_{\rm M}, \nonumber \\
{\bf R}_{1},{\bf R}_{2},...,{\bf R}_{\rm N},\alpha_{2},...,\alpha_{N})\},
 \label{BM_formula2_ind} 
\end{eqnarray}
where is assumed that the probability $P_{\mathrm{net}}$ depends
	on variables $q_{m}$,   $m=1,2,\ldots,M$ as well as ${\bf R}_{k}$ and $\alpha_{k}$, $k=1,2,\ldots,N$.

	In~\cite{BM_BM,BM_BM2,Kerner2014A}, it has been assumed that the probability  
		of traffic breakdown in the network
			\begin{eqnarray}
	\label{BM_formula3_sp} 
P_{\rm net}=1 - \prod^{N}_{k=1}(1-P^{({\rm B},k)}).
 \end{eqnarray}
  This means that we have assumed that
	   different traffic breakdowns occurring  at different network bottlenecks
		can be considered independent events.  
  In (\ref{BM_formula3_sp}), 
		  $P^{\rm (B,{\it k})}=P^{\rm (B,{\it k})}(q^{(k)}_{\rm sum}, \  \alpha_{k}, {\bf R}_{k})$ is the probability that during 
a time interval for observing traffic flow   $T_{\rm ob}$ 
 traffic breakdown 
occurs   at  bottleneck   $k$, where $k=1,2,\ldots,N$.
 
An analysis of traffic assignment in a   network with
 the BM principle (\ref{BM_formula2_ind}), (\ref{BM_formula3_sp}) made in~\cite{BM_BM}
permits  the  {\it distinct} assignment of network link inflow rates $q_{m}$ for the case 
of  relatively large values $Q$, when,  after the 
  application of   the BM principle
(\ref{BM_formula2_ind}), (\ref{BM_formula3_sp}), free flow remains in a metastable state
	 (\ref{C_min_q_in_larger}) at least at one of the network bottlenecks.
	In this case, although
	the application of the BM principle  (\ref{BM_formula2_ind}), (\ref{BM_formula3_sp}) 
	reduces the probability of traffic breakdown in the network to some minimum possible value
	$P_{\rm net}=P^{\rm (min)}_{\rm net}$,  
	this minimum value of the probability is    larger than zero:
					\begin{equation}
P_{\mathrm{net}}=P^{\rm (min)}_{\rm net}>0.
 \label{BM_network_larger_zero}
\end{equation}
No application of the BM principle
for the distinct assignment of network link inflow rates $q_{m}$ for the case 
				\begin{equation}
P_{\mathrm{net}}=0  
 \label{BM_network_zero}
\end{equation}
has been made in~\cite{BM_BM}.
The network throughput maximization approach
		resulting in formula (\ref{Cap_min_y_BM}) at $Z_{1}=Z$ as introduced in this article
			is the application of the BM principle
		(\ref{BM_formula2_ind}) for $\lq\lq$zero breakdown probability" (\ref{BM_network_zero}).
		The approach permits
	  the distinct assignment of network link inflow rates $q_{m}$ 
		under condition  (\ref{BM_network_zero}).

  Thus, there can be two different applications of the BM principle:
	
	(i) The application of the BM principle (\ref{BM_formula2_ind}), (\ref{BM_formula3_sp}) 
	related to the case
		(\ref{BM_network_larger_zero}). The    distinct  assignment of network link inflow rates $q_{m}$
		for this case has been made in~\cite{BM_BM}.

 (ii) The application of the BM principle
		(\ref{BM_formula2_ind}) for $\lq\lq$zero breakdown probability" (\ref{BM_network_zero}).
		This application made in   this article is the network throughput maximization approach.
		As shown in Sec.~\ref{Proc_BM_M},
		this approach resulting in conditions
		(\ref{Cap_min_y_BM})  at $Z_{1}=Z$ permits the distinct  assignment of network link inflow rates $q_{m}$
	for the case  (\ref{BM_network_zero}).

		When the value $Q$ is a relative small one, conditions (\ref{Cap_min_y_BM}) for the
		network throughput maximization   can be applied. 
		When the value $Q$ becomes a relative large one, 
		specifically, when conditions (\ref{C_min_q_sum_larger}) are satisfied,
		then the application of the BM principle
		(\ref{BM_formula2_ind}), (\ref{BM_formula3_sp}) of Ref.~\cite{BM_BM}
		to the distinct assignment of network link inflow rates $q_{m}$   related to the case
		(\ref{BM_network_larger_zero}) can be applied. 
		
	  We can conclude that the  
		network throughput maximization approach is applied to guarantee that condition (\ref{C_min_q_in_larger}) is
  satisfied at {\it none} of the network bottlenecks. This is possible
	as long as by the increase in the network inflow rate $Q$
	conditions (\ref{Cap_min_y_BM}) can be satisfied.
  When under subsequent increase in  the 
		total network inflow rate  $Q$ conditions (\ref{C_min_q_sum_larger}) are satisfied, 
  free flow is in a metastable state at some of the network bottlenecks.
	In this case, the  
		network throughput maximization approach cannot be applied any more. Instead, to minimize
		the probability of traffic breakdown, the application of the BM principle
		(\ref{BM_formula2_ind}), (\ref{BM_formula3_sp}) of Ref.~\cite{BM_BM}
		to the distinct assignment of network link inflow rates $q_{m}$   related to the case
		(\ref{BM_network_larger_zero}) can be applied. An analysis
		of the sequence of these different applications of  
 the BM principle   is out of scope of this paper; this 
 could be interesting task for future investigations.

\subsection{Conclusions \label{Con_S}}

	1.	We have revealed a network throughput maximization approach
	that
	ensures free flow conditions at which traffic breakdown cannot occur
	in a traffic or transportation network.

2. A physical   measure of a network  --   network
capacity is introduced that characterizes general features of the network with respect to the
 maximization of the network throughput. As long as
the total network inflow rate is smaller than
  the   network capacity no
traffic breakdown can
occur in the network.

3.	Based on the physics of the   network capacity, we have shown that
the classical Wardrop's UE or SO equilibrium
deteriorates basically the traffic system:
  Even when the total network inflow rate is smaller than  the   network capacity,
	the dynamic traffic assignment with the   Wardrop's equilibria
	leads to  the occurrence of traffic congestion in traffic and transportation networks.

\appendix
\section{Kerner-Klenov model for two-lane road with on-ramp bottleneck \label{App}}

In this Appendix, we present a discrete version of the Kerner-Klenov stochastic three-phase traffic flow model for single-lane road with on-ramp bottleneck~\cite{KKl2009A}.
 In the model, index $n$ corresponds 
to the discrete time $t_{\rm n}=\tau n, \ n=0,1,...$, 
$v_{n}$ is the vehicle speed at time step $n$, $a$ is the maximum acceleration,
$\tilde v_{n}$ is the vehicle speed  without  speed fluctuations, the lower index $\ell$  
marks variables related to the preceding vehicle, $v_{{\rm s}, n}$ is a safe speed at time step $n$,
$v_{\rm free}=v_{\rm free}(g_{n})$ is the maximum speed in free flow 
which is assumed to be a function of space gap $g_{n}$,
  $\xi_{n}$ describes   speed fluctuations;
 $v_{{\rm c},n}$ is a desired speed;
all vehicles have the same length $d$ that includes
the mean space gap between vehicles within a wide moving jam where the speed is  zero.
In the model, discretized space coordinate with a small
enough value of the discretization cell $\delta x$ is used. 
Consequently,  
  the vehicle speed and acceleration (deceleration) discretization intervals are $\delta v=\delta x/\tau$
 and   $\delta a=\delta v/\tau$, respectively, where time step $\tau=$ 1 s. Because in 
  the discrete model version   discrete (and dimensionless) values of space coordinate, speed and acceleration 
 are used, which are measured respectively in  values $\delta x$, $\delta v$ and  $\delta a$, 
and time is  measured in values of $\tau$,
   value $\tau$ in all formulas below is assumed to be the dimensionless value $\tau=1$. 
 In the model of an on-ramp bottleneck (Table~\ref{table1};
see explanations of model parameters in Fig.~16.2 (a)  of~\cite{KernerBook}),
superscripts    $+$   and  $-$  in variables, parameters, and functions 
denote the preceding vehicle and the trailing vehicle 
in the target lane during the lane changing on the main road or 
during the vehicle merging from the on-ramp lane.
 Initial and boundary conditions are the same as that explained in Sec.~16.3.9 of~\cite{KernerBook}. 
 Model parameters are presented in Tables~\ref{table_parameters} and~\ref{table_parameters_bottlenecks}.

\begin{table}
\caption{Discrete stochastic model~\cite{KKl2009A}}
\label{table_CA}
\begin{tabular}{|l|}
\hline
\multicolumn{1}{|c|}{
$v_{ n+1}=\max(0, \min({v_{\rm free}, \tilde v_{ n+1}+\xi_{ n}, v_{ n}+a
\tau, v_{{\rm s},n} })),$ 
}\\
\multicolumn{1}{|c|}{
$x_{n+1}= x_{n}+v_{n+1}\tau$,
}\\
\multicolumn{1}{|c|}{
$\tilde v_{n+1}=\min(v_{\rm free},  v_{{\rm s},n}, v_{{\rm c},n}),$
}\\
\multicolumn{1}{|c|}{
$v_{{\rm c},n}=\left\{\begin{array}{ll}
v_{ n}+\Delta_{ n} &  \textrm{at $g_{n} \leq G_{ n}$,} \\
v_{ n}+a_{ n}\tau &  \textrm{at $g_{n}> G_{ n}$}, \\
\end{array} \right.$
} \\
\multicolumn{1}{|c|}{
$\Delta_{ n}=\max(-b_{ n}\tau, \min(a_{ n}\tau, \ v_{ \ell,n}-v_{ n})),$
} \\
\multicolumn{1}{|c|}{
 $g_{n}=x_{\ell, n}-x_{n}-d$,
} \\
 $a$, $d$, and $\tau$ are constants, \\
$v_{\rm free}=v_{\rm free}(g_{n})$ is function of space gap $g_{n}$. \\
\hline
\end{tabular}
\end{table}
\vspace{1cm} 

\begin{table}
\caption{Functions in discrete stochastic model I: Stochastic time delay of acceleration and
deceleration}
\label{table_CA1}
\begin{center}
\begin{tabular}{|l|}
\hline
\multicolumn{1}{|c|}{$a_{n}=a  \Theta (P_{\rm 0}-r_{\rm 1})$, \
$b_{n}=a  \Theta (P_{\rm 1}-r_{\rm 1})$,} \\
\multicolumn{1}{|c|}{
$P_{\rm 0}=\left\{
\begin{array}{ll}
p_{\rm 0} & \textrm{if $S_{ n} \neq 1$} \\
1 &  \textrm{if $S_{ n}= 1$},
\end{array} \right.
\quad
P_{\rm 1}=\left\{
\begin{array}{ll}
p_{\rm 1} & \textrm{if $S_{ n}\neq -1$} \\
p_{\rm 2} &  \textrm{if $S_{ n}= -1$},
\end{array} \right.$
}\\
\multicolumn{1}{|c|}{
$S_{ n+1}=\left\{
\begin{array}{ll}
-1 &  \textrm{if $\tilde v_{ n+1}< v_{ n}$} \\
1 &  \textrm{if $\tilde v_{ n+1}> v_{ n}$} \\
0 &  \textrm{if $\tilde v_{ n+1}= v_{ n}$},
\end{array} \right.$
}\\
$r_{1}={\rm rand}(0,1)$, $\Theta (z) =0$ at $z<0$ and $\Theta (z) =1$ at $z\geq 0$; \\
$p_{\rm 0}=p_{\rm 0}(v_{n})$, $p_{\rm 2}=p_{\rm 2}(v_{n})$  are speed functions,
 $p_{\rm 1}$ is constant. \\
\hline
\end{tabular}
\end{center}
\end{table}
\vspace{1cm} 
  
\begin{table}
\caption{Functions in discrete stochastic model II: Model speed fluctuations}
\label{table_CA2}
\begin{center}
\begin{tabular}{|l|}
\hline
\multicolumn{1}{|c|}{
$\xi_{ n}=\left\{
\begin{array}{ll}
\xi_{\rm a} &  \textrm{if  $S_{ n+1}=1$} \\
- \xi_{\rm b} &  \textrm{if $S_{ n+1}=-1$} \\
\xi^{(0)} &  \textrm{if  $S_{ n+1}=0$},
\end{array} \right.$
}\\
\multicolumn{1}{|c|}{$\xi_{\rm a}=a^{(\rm a)} \tau \Theta (p_{\rm a}-r)$, \
$\xi_{\rm b}=a^{(\rm b)} \tau \Theta (p_{\rm b}-r)$,} \\
\multicolumn{1}{|c|}{
$\xi^{(0)}=a^{(0)}\tau \left\{
\begin{array}{ll}
-1 &  \textrm{if $r\leq p^{(0)}$} \\
1 &  \textrm{if $p^{(0)}< r \leq 2p^{(0)}$ and $v_{n}>0$} \\
0 &  \textrm{otherwise},
\end{array} \right.$
}\\
$r={\rm rand}(0,1)$;
 $p_{\rm a}$, $p_{\rm b}$, $p^{(0)}$, 
 $a^{(0)}$, 
$a^{(\rm a)}$, $ a^{(\rm b)}$
are constants, \\
\hline
\end{tabular}
\end{center}
\end{table}
\vspace{1cm}

\begin{table}
\caption{Functions in discrete stochastic model III: 
Maximum speed $v_{\rm free}$, synchronization gap $G_{n}$ and safe speed $v_{{\rm s},n}$}
\label{table_CA3}
\begin{center}
\begin{tabular}{|l|}
\hline
\multicolumn{1}{|c|}{
$v_{\rm free}=v_{\rm free}(g_{n}), $ 
} \\
\multicolumn{1}{|c|}{
$v_{\rm free}(g_{n})=\max(v^{(\rm min)}_{\rm free}, v^{(\rm max)}_{\rm free}(1-\kappa d/(d+g_{n}))), $ 
} \\
where   $v^{(\rm min)}_{\rm free}$, $v^{(\rm max)}_{\rm free}$ and $\kappa$ are constants.\\
\multicolumn{1}{|c|}{
$G_{n}=G(v_{n}, v_{\ell,n})$,
} \\
\multicolumn{1}{|c|}{
$G(u, w)=\max(0,  \lfloor k\tau u+  a^{-1}u(u-w) \rfloor),$
} \\
  $k>1$ is constant. \\
\multicolumn{1}{|c|}{
$v_{{\rm s},n}=
\min{(v^{\rm (safe)}_{ n},  g_{ n}/ \tau+ v^{\rm (a)}_{ \ell})},$
} \\
\multicolumn{1}{|c|}{
$v^{\rm (a)}_{\ell}=
\max(0, \min(v^{\rm (safe)}_{ \ell, n}, v_{ \ell,n}, g_{ \ell, n}/\tau)-a\tau),$
} \\
\multicolumn{1}{|c|}{
$v^{\rm (safe)}_{ n}=\lfloor v^{\rm (safe)} (g_{n}, \ v_{ \ell,n}) \rfloor,$ 
} \\
 $v^{\rm (safe)} (g_{n}, \ v_{ \ell,n}) $ is        taken  as that in~\cite{Kra10}, 
\\ which is a solution of  the
 Gipps's equation~\cite{Gipps10} \\
 \multicolumn{1}{|c|}{
$v^{\rm (safe)} \tau_{\rm safe} + X_{\rm d}(v^{\rm (safe)}) = g_{n}+X_{\rm d}(v_{\ell, n})$,
} \\
where   $\tau_{\rm safe}$
 is a safe time gap, \\
 \multicolumn{1}{|c|}{
$X_{\rm d} (u)=b \tau^{2} \bigg(\alpha \beta+\frac{\alpha(\alpha-1)}{2}\bigg)$,
} \\
\multicolumn{1}{|c|}{
$\alpha=\lfloor u/b\tau \rfloor$ and $\beta=u/b\tau-\alpha$ 
} \\
are the integer and  fractional parts  of $u/b\tau$, \\
respectively; 
$b$ is constant. \\
\hline
\end{tabular}
\end{center}
\end{table}
\vspace{1cm}

  \begin{table}
\caption{Lane changing rules from the right lane to the left lane ($R \rightarrow L$)
and from the left lane to the right lane ($L\rightarrow R$) and safety conditions
for lane changing 
}
\label{table_lane}
\begin{center}
\begin{tabular}{|l|}
\hline
\multicolumn{1}{|c|}{
$R \rightarrow L$: $v^{+}_{n} \geq v_{\ell, n}+\delta_{1}$   and $v_{n}\geq v_{\ell, n}$,
}\\
\multicolumn{1}{|c|}{
$L \rightarrow R$: $v^{+}_{n} > v_{\ell, n}+\delta_{1}$ or $v^{+}_{n}>v_{n}+\delta_{1}$.
}\\
\multicolumn{1}{|c|}{Safety conditions:} \\
\multicolumn{1}{|c|}{
$g^{+}_{n} >\min(v_{n}\tau, \ G^{+}_{n})$,
}\\
\multicolumn{1}{|c|}{  
$g^{-}_{n} >\min(v^{-}_{n}\tau, \ G^{-}_{n})$, 
}\\
\multicolumn{1}{|c|}{
$G^{+}_{n}=G( v_{n}, v^{+}_{n})$,
 $G^{-}_{n}=G(v^{-}_{n}, v_{n})$,
 }\\
  $G(u, w)$
is given in Table~\ref{table_CA3}; \\
lane changing occurs with probability $p_{\rm c}$. \\
\hline
\end{tabular}
\end{center}
\end{table}
\vspace{1cm}

\begin{table}
\caption{Models of vehicle merging at on-ramp bottleneck
that occurs when   a safety rule ($\ast$) {\it or} a safety rule  ($\ast \ast$) is satisfied 
}
\label{table1}
\begin{center}
\begin{tabular}{|l|}
\hline
\multicolumn{1}{|c|}{Safety rule ($\ast$):}\\
\multicolumn{1}{|c|}{
$\begin{array}{ll}
g^{+}_{n} >\min(\hat  v_{n}\tau , \ G(\hat  v_{n}, v^{+}_{n})), \\
g^{-}_{n} >\min(v^{-}_{n}\tau, \ G(v^{-}_{n},\hat  v_{n})),
\end{array} $
}\\
\multicolumn{1}{|c|}{
$\hat v_{n}=\min(v^{+}_{n},  \ v_{n}+\Delta v^{(1)}_{r}),$
} \\
 $\Delta v^{(1)}_{r}>0$ is constant.\\
\hline
\multicolumn{1}{|c|}{Safety rule ($\ast \ast$):}\\
\multicolumn{1}{|c|}{
$x^{+}_{n}-x^{-}_{n}-d > \lfloor  \lambda_{\rm b} v^{+}_{n} +d \rfloor,$
}\\
\multicolumn{1}{|c|}{
$\begin{array}{ll}
x_{n-1}< x^{\rm (m)}_{n-1} \  \textrm{and} \
 x_{n} \geq x^{\rm (m)}_{n} \\
\ \textrm{or} \
x_{n-1} \geq x^{\rm (m)}_{n-1} \  \textrm{and} \
 x_{n} < x^{\rm (m)}_{n},
\end{array}$
}\\
\multicolumn{1}{|c|}{
$x^{\rm (m)}_{n}=\lfloor (x^{+}_{n}+x^{-}_{n})/2 \rfloor,$
}\\
$\lambda_{\rm b}$ is constant. \\
\hline
\multicolumn{1}{|c|}{Parameters after vehicle merging:}\\
\multicolumn{1}{|c|}{$v_{n}=\hat v_{n},$}\\
\multicolumn{1}{|c|}{under the rule ($\ast $): $x_{n}$  maintains the
same,}\\
\multicolumn{1}{|c|}{under the rule ($\ast \ast$): $x_{n} = x^{\rm
(m)}_{n}$.}\\
\hline
\multicolumn{1}{|c|}{Speed adaptation before vehicle merging}\\
\multicolumn{1}{|c|}{
$v_{{\rm c},n}=\left\{\begin{array}{ll}
v_{ n}+\Delta^{+}_{ n} &  \textrm{at $g^{+}_{n} \leq G(v_{n}, \hat
v^{+}_{n})$,} \\
v_{ n}+a_{ n}\tau &  \textrm{at $g^{+}_{n}>G( v_{n}, \hat
v^{+}_{n})$}, \\
\end{array}\right. $
}\\
\multicolumn{1}{|c|}{
$\Delta^{+}_{ n}=\max(-b_{ n}\tau, \min(a_{ n}\tau, \ \hat v^{+}_{n}-v_{
n})),$
}\\
\multicolumn{1}{|c|}{
$\hat v^{+}_{n}=\max(0, \min(v^{\rm (max)}_{\rm free}, 
\  v^{+}_{n}+\Delta
v^{(2)}_{r})),$
}\\
$\Delta v^{(2)}_{r}$ is  constant. \\
\hline
\end{tabular}
\end{center}
\end{table}
\vspace{1cm}

\begin{table}
\caption{Model parameters: Vehicle motion in road lane}
\label{table_parameters}
\begin{center}
\begin{tabular}{|l|}
\hline
$\tau_{\rm safe}   = \tau=$ 1, $d = 7.5 \  \rm m/\delta x$, \\
$\delta x=$ 0.01 m, $\delta v= 0.01 \  {\rm ms^{-1}}$, $\delta a= 0.01 \  {\rm ms^{-2}}$, \\
$v^{(\rm min)}_{\rm free}=25 \ {\rm ms^{-1}}/\delta v$, 
$v^{(\rm max)}_{\rm free}=41.67 \ {\rm ms^{-1}}/\delta v$, \\
$\kappa=1.73$, 
$b = 1 \ {\rm ms^{-2}}/\delta a$, $a=$ 0.5 ${\rm ms^{-2}}/\delta a$, \\
$k=$ 3, $p_{1}=$ 0.3,  $p_{b}=   0.1$,  
$p_{a}=   0.17$,
 $p^{(0)}= 0.005$, \\
$p_{\rm 2}(v_{n})=0.48+ 0.32\Theta{( v_{n}-v_{21})}$, \\
$v_{01} = 10 \ {\rm ms^{-1}}/\delta v$, $v_{21} = 15 \ {\rm ms^{-1}}/\delta v$, \\
$p_{\rm 0}(v_{n})=0.575+ 0.125\min{(1, v_{n}/v_{01})}$, \\
  $a^{(0)}= 0.2a$,   $a^{(\rm a)}= a$,   \\  
 $a^{(\rm b)}(v_{n})=0.2a+
  0.8a\max(0, \min(1, (v_{22}-v_{n})/\Delta v_{22})$, \\
  $v_{22} = 12.5 \ {\rm ms^{-1}}/\delta v$,  
  $\Delta v_{22} = 2.778 \ {\rm ms^{-1}}/\delta v$. \\
   \hline
\end{tabular}
\end{center}
\end{table}
\vspace{1cm}

\begin{table}
\caption{Model parameters: Lane changing}
\label{table_parameters_lane_changing}
\begin{center}
\begin{tabular}{|l|}
\hline  
$\delta_{1}=1$  
  ${\rm ms^{-1}}/\delta v$,  
  $p_{\rm c}=0.2 $.  \\
\hline
\end{tabular}
\end{center}
\end{table}
\vspace{1cm}

\begin{table}
\caption{Parameters of model of on-ramp bottleneck}
\label{table_parameters_bottlenecks}
\begin{center}
\begin{tabular}{|l|}
\hline
$\lambda_{\rm b}=$ 0.75, 
   $v_{\rm free \ on}=22.2 \ {\rm ms^{-1}}/\delta v$,  \\
   $\Delta v^{\rm (2)}_{\rm r}=$ 5 \  ${\rm ms^{-1}}/\delta v$, 
   $L_{\rm r}=1 \ {\rm km}/\delta x$,  \\ $\Delta v^{\rm (1)}_{\rm r}=10 \ {\rm ms^{-1}}/\delta v$,  
   $L_{\rm m}=$ 0.3 \    ${\rm km}/\delta x$. \\
   \hline
\end{tabular}
\end{center}
\end{table}
\vspace{1cm}

 {\bf Acknowledgment:}
I thank Sergey Klenov for discussions and help in simulations.


\begin{thebibliography}{8.}
\addcontentsline{toc}{section}{References} 

\bibitem{Wardrop}
 J.G.	Wardrop,  in: Proc. of Inst. of Civil Eng. II.   1   (1952)  325--378.

\bibitem {Sheffi1984}  
Y. Sheffi, 
Urban transportation networks: Equilibrium analysis with mathematical programming
methods. New Jersey: Prentice-Hall, 1984.


 

\bibitem {BellIida19897}
M.G.H. Bell, Y. Iida, Transportation network analysis, 
John Wiley {\&} Sons, Incorporated, Hoboken, NJ 07030-6000, USA,
1997.

\bibitem {Peeta2001A}
 S. Peeta, A.K. Ziliaskopoulos,  
Networks and Spatial Economics  {\bf 1} 233--265  (2001).

\bibitem {Mahmassani2001A}
 H.S. Mahmassani,  
Networks and Spatial Economics  {\bf 1} 267--292  (2001).

 \bibitem {Rakha2009}

 H.  Rakha,  A.  Tawfik,  
 in {\it Encyclopedia of Complexity and System Science}, 
ed. by R.A. Meyers. (Springer, Berlin, 2009), pp. 9429--9470.

\bibitem{DaganzoSheffi1977}
C.F. 	Daganzo, Y. Sheffi,  Transp. Sci. 
 	 {\bf 11} 253--274  (1977).
	
	\bibitem{Merchant1978A}
D.K. Merchant, G.L. Nemhauser, Transp. Sci.  {\bf 12} 
 187--199 (1978). 

\bibitem{Merchant1978B}
D.K. Merchant, G.L. Nemhauser, Transp. Sci.  {\bf 12} 
    (1978).

\bibitem{Bell1992A}
M.G.H. Bell, Transp. Res. B {\bf  26} 
 303--313 (1992).

\bibitem{Bell2002A}
M.G.H. Bell, Transp. Res. B
     {\bf 36}
671--681 (2002).

\bibitem{Mahmassani1987A}

H.S.  Mahmassani,   G.L.  Chang.  Transp. Sci.  {\bf 21} 89--99  (1987).

\bibitem{Friesz2013A}


 T.L. Friesz,   K. Han,  P.A. Neto, A. Meimand, T. Yao,  Transp. Res.   B
 {\bf 47} 102--126    (2013).

\bibitem{Ben-Akiva2015A}

L. Lu, Y. Xu, C. Antoniou,  M. Ben-Akiva
Transport. Res.  C {\bf 51} 149--166 (2015).


 
 \bibitem{Wahle}
 J. Wahle, A.L.C. Bazzan, F. Klugl, M. Schreckenberg, Physica A {\bf 287} 669 (2000).

\bibitem{Davis1}
 L.C. Davis,  Physica A {\bf 388} 4459  (2009).

\bibitem{Davis2}
 L.C. Davis,  Physica A  {\bf 389} 3588  (2010).

\bibitem{Wang2005A}
Wen-Xu Wang,
Bing-Hong Wang, Wen-Chen Zheng, Chuan-Yang Yin, and Tao Zhou
Phys. Rev. E {\bf 72}
066702 (2005).

\bibitem{Claes2011A}
R. Claes, T. Holvoet, and D. Weyns,
IEEE Trans. on ITS  {\bf 12} 364--373 (2011).

\bibitem{Donga2010A}
Chuan-Fei Donga,  Xu Mab,   Bing-Hong Wang,
Phys. Lett. A
{\bf 374},  1326--1331 (2010).

\bibitem {Sopasakis2006}
A. Sopasakis, M.A. Katsoulakis,
SIAM J.
Appl. Math. {\bf 66}, 921--944 (2006).

\bibitem {Hauck2014}
Cory Hauck, Yi Sun, and Ilya Timofeyev, Stoch. Dyn. {\bf 14}, 1350022 (2014).

\bibitem {Sun2014A}
Yi Sun and Ilya Timofeyev
Phys. Rev. E {\bf 89}, 052810 (2014).

  \bibitem {Kerner_Review} 
B.S. Kerner,  Physica A  {\bf 392}  5261--5282  (2013).

 \bibitem{MiniReview2}

B.S. Kerner,  Physica A   
{\bf 450} 700--747  (2016).

 	\bibitem{KernerBook}
B.S. Kerner, {\em The Physics of Traffic} 
(Springer, Berlin,   2004).

 \bibitem{KernerBook2}
B.S. Kerner, {\it Introduction to Modern Traffic Flow Theory and Control.}
	 (Springer, Berlin,   2009).
	
	
			\bibitem{BM_BM}
   	B.S. Kerner, J. Phys. A: Math. Theor.  {\bf 44} 092001  (2011).
 

\bibitem{BM_BM2}
   	B.S. Kerner,  
		Phys. Rev. E {\bf 84}, 045102(R) (2011).
	
 \bibitem{ElefteriadouBook2014}
  
 L.  Elefteriadou,  An Introduction to Traffic Flow Theory.
(Springer, Berlin 2014).

\bibitem {May1990}

A.D. May,  
 Traffic Flow Fundamentals. 
 (Prentice-Hall, Inc., New Jersey, 1990).
		
		

\bibitem{Reviews2}
D. Helbing,
Rev. Mod. Phys. {\bf 73}   1067--1141 (2001);
D. Chowdhury, L. Santen,  A. Schadschneider, 
Phys. Rep. {\bf 329}   199 (2000);
M. Treiber, A. Kesting, Traffic Flow Dynamics (Springer, Berlin, 2013).
Springer Optimization and Its Applications, Vol. 84, Springer, Berlin (2014).
		
  
		

		
  
		

\bibitem{ControlParemeters}
 Examples of   possible control parameters of network bottlenecks are
 the on-ramp inflow
  $q_{\rm on,{\it k}}$ for the
 case when bottleneck $k$ is an on-ramp bottleneck  as well as the flow rate of vehicles $q_{\rm off,{\it k}}$ 
leaving the main road to an off-ramp
for the
 case when bottleneck $k$ is an off-ramp bottleneck.





\bibitem{DiffVehicles}
For example,
different assignment can be done
 for long vehicles (trucks) and passenger vehicles as well as for usual vehicles and electric  vehicles.

\bibitem{Equil}

 Usually, the network inflow   rate $Q(t)$ changes in a network 
	over time very slowly in comparison with
	any characteristic times of dynamic 
	traffic effects at network bottlenecks under free flow conditions in
	the network. For this reason, as usually assumed in
  theories of traffic and transportation networks
	(see, e.g.,~\cite{Wardrop,Sheffi1984,BellIida19897,Peeta2001A,Mahmassani2001A,Rakha2009}),
	at any given $Q$  free flow distribution in the network  
	can be considered as a steady   state (steady-state analysis of
	traffic and transportation networks). This means that
 $Q=Q_{\rm out}$,
		where $Q_{\rm out}(t)=\sum^{J}_{j=1}{q^{\rm (d)}_{j}(t)}$ is the total network outflow rate. 
		
 
		




 \bibitem{Cal_C}
 To explain the term {\it maximum total network inflow rate} in the definition of the 
  network capacity $C_{\rm net}$
in more details,
we note that in some cases the minimum capacity $C_{\rm min}$ of a network
 bottleneck can be a function of flow rates $q_{m}$.
For example, for an   on-ramp bottleneck 
$C_{\rm min}$ can be a function of
  the on-ramp inflow rate   $q_{\rm on}$.
For this reason, to find   $C_{\rm net}$,
  the assignment of the flow rates $q_{m}$
as well as    control of bottleneck parameters should ensure
the maximum total network inflow rate $Q$ at which conditions (\ref{C_min_q_sum_equal})  
are satisfied.
A development of a general procedure for the calculation of  
the  network capacity for    traffic networks
is out of scope of this paper. Two examples 
of   calculations $C_{\rm net}$ for
simple network models are presented in Sec.~\ref{Numerical_S}.

\bibitem{Cal_C2}
Note that   the   network capacity
$C_{\rm net}$   depends 
on the network
 inflow rates $q^{\rm (o)}_{ij}(t)$ and the  set of the alternative routes, which  
are assumed to be given.  For the same day,
 there can be also different
values   
$C_{\rm net}$   at different time instants.

 	\bibitem{City}	
  In city traffic,
there is  a hypothetical case of $\lq\lq$red wave", when all
vehicles  approach  a traffic signal during the red signal phase {\it only}. In this case,
the definition of the   network capacity $C_{\rm net}$
through conditions (\ref{C_min_q_sum_equal})  remains. However, when at
one of the network bottlenecks due to the    signal  
$\lq\lq$red wave"   is realized, then the minimum capacity of the signal
$C_{\rm min}$   is equal to  the classical signal capacity $C_{\rm cl}$~\cite{Kerner2014A,Kerner2014B}.
Therefore, in (\ref{C_min_q_sum_equal})  the value of $C_{\rm min}$
for this signal should be replaced by $C_{\rm cl}$. Moreover, 
when   the signal is one of the
network bottlenecks $k=k^{(2)}_{w}$ in (\ref{C_min_q_sum_larger}), 
  then over-saturated (congested) traffic does occur
at this signal.  
To explain this, we note that in the case of
 $\lq\lq$red wave"
the classical theory of traffic at the signal is a special case of 
the three-phase theory~\cite{Kerner2014A,Kerner2014B}: 
When the average arrival flow rate (flow rate at a bottleneck due to the signal) exceeds
  $C_{\rm min}=C_{\rm cl}$, 
  then  traffic breakdown, i.e.,
the transition from under-saturated traffic to over-saturated
(congested) traffic occurs
at the signal without time delay.  
The network capacity $C_{\rm net}$ follows also  
from the above application of the network throughput maximization  approach, if
in all formulas of Sec.~\ref{Proc_BM_C_min}
 we replace $C^{(k)}_{\rm min}$   by  $C^{(k)}_{\rm cl}$. In particular,
  $C_{\rm net}$ is determined from conditions
 (\ref{C_min_q_sum_equal}) as follows: 
\begin{eqnarray}
 \label{C_min_q_sum_equal_cl} 
  q^{(k)}_{\rm sum} = C^{(k)}_{\rm cl}  \  {\rm for}   	
\ k=k^{(1)}_{z},  	 
\\  q^{(k)}_{\rm sum} < C^{(k)}_{\rm cl} \ {\rm for} 
 \ k \ne k^{(1)}_{z},
 \nonumber	
	\\ 	  z=1,2,\ldots,Z; \ Z\geq 1, \  Z\leq N; \ k=1,2,\ldots,N, \nonumber 
     \end{eqnarray}
where bottlenecks $k=k^{(1)}_{z}$ and value $Z$ are found
in accordance with  the constrain $\lq\lq$alternative routes" as described in Sec.~\ref{Proc_BM_M}.
Formula (\ref{C_min_q_sum_equal_cl})
is also applicable in the framework of the classical traffic flow theories in which is assumed that there is 
a particular value of capacity   for any network bottleneck.
If we denote   capacity of free flow at network bottleneck $k$ by $C^{(k)}_{\rm cl}$,
formula (\ref{C_min_q_sum_equal_cl}) remains for any classical traffic flow theory.
In other words, the application of the network throughput maximization approach
 leading to formula (\ref{C_min_q_sum_equal_cl})
does not depend on a traffic flow theory applied for the explanation of
 the physical nature of  highway capacity.
 In particular, 
the measure $\lq\lq$network capacity" has the same sense
in the classical theory as that in the three-phase traffic theory.
In a steady state analysis of a network~\cite{Equil},
condition
 $Q \rightarrow C_{\rm net}$ provides 
the maximum possible network throughput at which free flow conditions
are ensured at which traffic breakdown cannot occur  
  in the whole network.




\bibitem{Kerner2014A}

B.S. Kerner, Physica A {\bf 397},  76--110 (2014).


\bibitem{Kerner2014B}

B. S. Kerner, S.L. Klenov, M. Schreckenberg,  J.
of Stat. Mech.,     P03001 (2014).

\bibitem{Q_change}
	As the total flow rate  $Q$ in (\ref{Cap_min_net_BM_larger})
		is subsequently  increased, the number $W$ of bottlenecks
  for which
		conditions (\ref{C_min_q_sum_larger}) are satisfied, can   change.

\bibitem{Meta_Ward}

In numerical simulations of the application of the Wardrop's equilibria
to dynamic traffic assignment in two simple models of networks (Sec.~\ref{Numerical_S}),
the metastability of free flow  with respect to an F$\rightarrow$S transition
(traffic breakdown) at network bottlenecks has been   taken into account
	in travel time costs through the use of the Kerner-Klenov microscopic stochastic
	three-phase model.
	
	\bibitem{TwoRoute}
	Two-route models are often used for   studies
	of other different  traffic phenomena
		(e.g.,~\cite{Wahle,Davis1,Davis2}).

\bibitem{Num_Sim}
Numerical simulations of more complex networks are out of scope of this article. 
 Simulations for specific networks could 
be interesting task for further investigations. 
   
 
  
 \bibitem{KKl}

B.S. Kerner,  S.L. Klenov, 
 J. Phys. A: Math. Gen.  {\bf 35}, L31--L43 (2002);
  B.S. Kerner,  S.L. Klenov,
Phys. Rev. E  {\bf 68} 036130 (2003);
B.S. Kerner and S.L. Klenov. J. Phys. A: Math. Gen. {\bf 37} 8753--8788  (2004).

   

 \bibitem{KKl2009A}
 B.S. Kerner,  S.L. Klenov, Phys. Rev. E
{\bf 80} 056101 (2009).

\bibitem{UE_T}
	
	The dynamic traffic  assignment of
   the  link inflow  rates 
$q_{m}$  ($m=1, 2$ in Fig.~\ref{BM_Network_Model2_new} (a)
	and $m=1, 2, 3$ in Fig.~\ref{BM_Network_Model2_new} (c)) under time-dependent flow rate $q^{\rm (o)}(t)$
		is performed with the use of
		standard methods~\cite{Sheffi1984,BellIida19897,Peeta2001A,Mahmassani2001A,Rakha2009}:
		For some flow rates $q^{\rm (o)}$, values $q_{m}$ have been found in accordance with the
		Wardrop's UE
		(points in Figs.~\ref{UE_ANCONA_2} (b) and~\ref{UE_ANCONA_3} (b)).
		Simulations show that these points are well fitted with lines
		given in captions to Figs.~\ref{UE_ANCONA_2}  and~\ref{UE_ANCONA_3}.
		These lines have further been used for calculations of $q_{m}$.
	
	\bibitem{ANCONA}

We have proven 
different control methods to decrease congestion in the network occurring 
due to dynamic traffic assignment with Wardrop's UE or SO (as above-explained it is not possible to
avoid congestion fully). Some of the best results shown in
Figs.~\ref{UE_ANCONA_5900_short}--\ref{UE_ANCONA_3_2} are achieved with  
$\lq\lq$congested pattern control method"~\cite{KernerBook}, in which
bottleneck control starts {\it only} after    traffic breakdown (F$\rightarrow$S transition)
 is registered on the shortest route (route 1) through the use of
 a road detector installed  500 m upstream of bottleneck location $x_{\rm on}=$ 15 km. 
An  F$\rightarrow$S transition is  registered  when the speed at the detector $v\leq 80$ km/h
(for two-route network) and $v\leq 70$ km/h (for three route network)
during 3 min; a return
S$\rightarrow$F transition is  registered  when the speed  
at the detector $v\geq 92$ km/h during 2 min. 
	After the F$\rightarrow$S transition 
		has been registered on bottleneck 1,
the flow rate   on   route 1 decreases on a value $\Delta q$ 
and the flow rate   on   route 2 increases on the same value $\Delta q$
(Figs.~\ref{UE_ANCONA_5900_short} and~\ref{UE_ANCONA_2}).
For three-route network (Fig.~\ref{UE_ANCONA_3_2}), 
the  increase in the flow rate $\Delta q$ has been divided 
(in accordance with Eq.~(\ref{UE_eq3}) for free flow conditions)
between two other routes (routes 2 and 3).
After a return S$\rightarrow$F transition 
		has been registered at route 1,
		the flow rates $q_{k}$   
  return to their initial values
  (at $t=0$).  
	
	
		
	
	 \bibitem{SO}
	Qualitative the same results as shown in  Figs.~\ref{UE_Prob_5900_short}
	and~\ref{UE_ANCONA_5900_short}
	have been derived for Wardrop's SO at
	$Q=7300<C^{\rm (net)}_{\rm min}=7740$ vehicles/h.

\bibitem{Hemmerle2015}
P. Hemmerle, M. Koller, H. Rehborn, B. S. Kerner, M. Schreckenberg,  
 IET Intell. Transp. Syst.,   1--8 ( 2015),  doi: 10.1049/iet-its.2015.0014. 


  \bibitem{Kerner_CO2}
	  While   working at the Daimler Company,
 the author was lucky to take part in the development of systems for traffic information
used in vehicle routing systems,
which are on the market.
 

	 

 \bibitem {Gipps10} 
 P.G. Gipps,
Trans. Res. B  15  (1981) 105--111.

 
 

 \bibitem{Kra10}

 S.  Krau{\ss},  P.   Wagner,  C. Gawron,
Phys. Rev. E    55  (1997)  5597--5602.

 
  
      
\end{thebibliography}
\end{document}